\documentclass[11pt,a4paper]{article}
\usepackage{jheppub}
\usepackage{dsfont}
\usepackage[vcentermath]{youngtab}

\usepackage{amsmath, amssymb}								
\usepackage{amsmath, amssymb,wasysym}
\usepackage{mathrsfs}

\newcommand{\be}{ \begin{equation}}
\newcommand{\ee}{\end{equation}}
\newcommand{\bea}[1]{\begin{eqnarray}\label{#1} }
\newcommand{\eea}{\end{eqnarray}}

\newcommand{\hs}[1]{\mbox{hs$[#1]$}}
\newcommand{\w}[1]{\mbox{$\mathcal{W}_\infty[#1]$}}

\newcommand{\mb}{\mathbb}
\newcommand{\super}[1]{\mb V^{(\frac{#1}{2})}}
\newcommand{\extrasuper}[2]{\mb V^{(\frac{#1}{2}), #2}}

\newcommand{\extracompsuper}[2]{\mb V^{(#1), #2}}
\newcommand{\coupling}[3]{\mb D^{#1\,#2}_{#3}}
\newcommand{\wt}{\widetilde}

\title{$\mathcal{N}=1$ extension of  minimal model holography}

\author[a]{Matteo Beccaria,}
\author[b]{Constantin Candu,}
\author[b]{Matthias R. Gaberdiel,}
\author[b]{Michael Groher}

\affiliation[a]{Dipartimento di Matematica e Fisica ìEnnio De Giorgiî,\\
Universitaí del Salento \& INFN, Via Arnesano, 73100 Lecce, 
Italy}
\affiliation[b]{Institut f\"ur Theoretische Physik,\\ 
ETH Zurich, CH-8093 Z\"urich, Switzerland}

\emailAdd{matteo.beccaria@le.infn.it}
\emailAdd{canduc@itp.phys.ethz.ch}
\emailAdd{gaberdiel@itp.phys.ethz.ch}
\emailAdd{mgroher@student.ethz.ch}

\abstract{
The CFT dual of the higher spin theory with minimal ${\cal N}=1$ spectrum is determined. 
Unlike previous examples of minimal model holography, there is no free parameter beyond
the central charge, and the CFT can be described in terms of a non-diagonal modular invariant
of the bosonic theory at the special value of the 't~Hooft parameter $\lambda=\tfrac{1}{2}$. As 
evidence in favour of the duality we show that 
the symmetry algebras as well as the partition functions agree between the two descriptions.}

\allowdisplaybreaks
\begin{document}

\maketitle
\flushbottom

\section{Introduction}

Recently,  dualities relating higher spin theories on AdS spaces \cite{Vasiliev:2003ev}
to vector-like conformal field theories have attracted some attention. The original idea was
already suggested some time ago \cite{Sundborg:2000wp,Witten,Mikhailov:2002bp} and
first concrete proposals were made soon thereafter \cite{Klebanov:2002ja,Sezgin:2003pt}, but
it was only through the work of Giombi \& Yin \cite{Giombi:2009wh,Giombi:2010vg} that 
compelling evidence was obtained. Dualities of this kind are very interesting because
they hold the promise of offering insights into the conceptual underpinning of the AdS/CFT
correspondence.

More recently, a lower dimensional version, relating higher spin theories on AdS$_3$ 
\cite{Prokushkin:1998bq,Prokushkin:1998vn} to the large $N$ limit of some 2d minimal model
CFTs was proposed \cite{Gaberdiel:2010pz}. These incarnations are interesting since
higher spin theories in 3d can be much more easily described in terms of a Chern-Simons formulation.
At the same time, 2d minimal model CFTs are under very good analytical control, and thus the
correspondence can be analysed and tested in quite some detail, see e.g.\ 
\cite{Henneaux:2010xg,Campoleoni:2010zq,Gaberdiel:2011wb,Gaberdiel:2012ku} for the matching
of the symmetries; \cite{Gaberdiel:2011zw,Castro:2011iw,Perlmutter:2012ds} for the comparison of 
the spectrum; and \cite{Papadodimas:2011pf,Chang:2011mz,Chang:2011vk,Ahn:2011by,Ammon:2011ua} for 
the analysis of correlation functions.
Finally, these models evade the Maldacena-Zhiboedov theorem \cite{Maldacena:2011jn,Maldacena:2012sf}
that implies that higher dimensional
theories with an unbroken higher spin symmetry and finitely many degrees of freedom are
necessarily free. For a review of this circle of ideas see \cite{Gaberdiel:2012uj}.

The original proposal of \cite{Gaberdiel:2010pz} has been generalised in a variety of directions: 
to the case with orthogonal gauge groups \cite{Ahn:2011pv,Gaberdiel:2011nt}, the situation with
${\cal N}=2$ supersymmetry \cite{Creutzig:2011fe} (see also \cite{supersymmetric}), and more 
recently to the case with ${\cal N}=1$ supersymmetry \cite{Creutzig:2012ar} (see also \cite{CV}). In this paper we 
give evidence for a different ${\cal N}=1$ supersymmetric duality, for which the spin content
of the higher spin gauge fields is minimal, i.e.\ it consists of a single higher spin field for 
each half-integer spin $s\geq \tfrac{3}{2}$. Unlike the previous examples, there
is no additional free parameter (except for the central charge that is proportional to the radius of 
the AdS space) in this case: this can be seen from the AdS point of view where the 
corresponding higher spin algebra is the ${\rm shs}(1|2)$ algebra of \cite{fradkin} that does not
have a deformation parameter; it also follows from the analysis of the most general $s{\cal W}_{\infty}$
algebra with the above spin content for which we have found (see Section~\ref{sec:sWalgebra}) 
that the Jacobi identities do not allow for a free coupling constant. In fact, as we shall explain in 
Section~\ref{sec:bosex}, the relevant CFT can be described as the ${\cal N}=1$ supersymmetric
extension of the bosonic theory of \cite{Gaberdiel:2010pz} that exists provided that the level 
of the numerator $\mathfrak{su}(N)_k$ algebra equals $k=N$. From this point of view, the
${\cal N}=1$ supersymmetric theory then corresponds to a `non-diagonal' modular invariant. Our
duality is therefore the first interesting example where one of the non-diagonal coset modular invariants
has been identified with a dual higher spin theory.
\medskip

The paper is organised as follows. In Section~\ref{sec:bosre} we review some of the salient features
of the bosonic duality of \cite{Gaberdiel:2010pz} that will be relevant in the following. Section~\ref{sec:bosex}
explains how this theory can be extended for $k=N$ to an ${\cal N}=1$ superconformal field theory. In particular,
we calculate the extended characters and give an explicit formula for the full ${\cal N}=1$ superconformal
partition function, see eq.~(\ref{part2}). In Section~\ref{sec:sWalgebra} we then analyse the most general
${\cal N}=1$ superconformal $s{\cal W}$-algebra with the given spectum, and show that it does not possess 
any free parameter (except for the central charge). We also show that it contains the bosonic 
${\cal W}_{\infty}[\tfrac{1}{2}]$ algebra as a subalgebra (see Section~\ref{sec:bossub}), as expected from
the above construction. Section~\ref{sec:AdShs} describes the dual higher spin theory, and shows that
the asymptotic symmetry algebra of the Chern-Simons theory based on ${\rm shs}(1|2)$ agrees indeed
with the `wedge' algebra of our $s{\cal W}$-algebra. Finally, we explain how the full spectrum of the
coset CFT in the 't~Hooft limit can be accounted for by adding to the  higher spin fields a complex
${\cal N}=1$ matter multiplet, see Section~\ref{sec:part}. Section~\ref{sec:concl} contains a brief conclusion, and there are
three appendices where some of the more technical material is explained.

\section{Review of bosonic minimal model holography}\label{sec:bosre}

Let us start by reviewing the bosonic duality of \cite{Gaberdiel:2010pz}, see \cite{Gaberdiel:2012uj} for a recent review. 
The bosonic higher spin theory on AdS$_3$ 
based on the Lie algebra $\hs{\lambda}$ \cite{Prokushkin:1998bq,Prokushkin:1998vn} is conjectured to be dual
to the 't~Hooft like large $N$ limit of the cosets
\begin{equation} \label{bosoniccoset} 
\frac{\mathfrak{su}(N)_k \oplus \mathfrak{su}(N)_1}{\mathfrak{su}(N)_{k+1}} 
\end{equation}
with central charge 
\begin{equation} \label{centralcharge} 
c=(N-1) \left( 1-\frac{N(N+1)}{(N+k)(N+k+1)} \right)\ ,
\end{equation}
where $\lambda$ is identified with the 't~Hooft coupling 
\be
\lambda = \frac{N}{N+k} 
\ee
that is held fixed in the large $N,k$ limit. In particular, the symmetries of the coset define a $\w{\mu}$ algebra that is generated
by the stress energy tensor, together with one Virasoro primary field of each integer spin $s\geq 3$. For a given value of 
the central charge $c$, the $\w{\mu}$ algebras corresponding to three (generically different) values of $\mu$ describe isomorphic
${\cal W}$-algebras \cite{Gaberdiel:2012ku}, and this `triality' of relations explains, in particular, why the quantisation of the 
asymptotic symmetry algebra of the higher spin theory (that was first determined classically in 
\cite{Henneaux:2010xg,Campoleoni:2010zq,Gaberdiel:2011wb}) agrees with that of the dual coset. 

In addition to the vacuum representation (that just describes the $\w{\lambda}$ algebra), the 
coset theory contains a number of irreducible representations that are labelled by the pairs $(\Lambda_+;\Lambda_-)$, 
where $\Lambda_+$ and $\Lambda_-$ are representation of $\mathfrak{su}(N)_k$ and $\mathfrak{su}(N)_{k+1}$, 
respectively. These degrees of freedom correspond, on the higher spin side, to those of a complex scalar field with mass
\be
M^2 = - (1-\lambda^2) \  ,
\ee
as well as a number of classical solutions \cite{Castro:2011iw,Gaberdiel:2012ku,Perlmutter:2012ds}, see
also \cite{Jevicki:2013kma,Chang:2013izp} for a somewhat different interpretation.
The spectrum of the coset theories is taken to be given by the `A-type modular invariant' partition function, i.e.\ it is of the form
\be
{\cal H} = \bigoplus_{(\Lambda_+,\Lambda_-)} (\Lambda_+;\Lambda_-) \otimes \overline{(\Lambda^\ast_+;\Lambda^\ast_-)} \ , 
\ee
where the sum runs over all inequivalent coset representations, and $\Lambda^\ast$ denotes the 
representation conjugate to $\Lambda$. The states in the representations $(\Lambda_+;0)$ correspond
to the excitations of the  complex scalar field, while the remaining states account for the classical solutions (that are labelled
by $(0;\Lambda_-)$), as well as its scalar excitations. In the strict 't~Hooft limit, some of the latter states decouple, 
and the resulting partition function agrees precisely with the 1-loop thermal partition function of the higher spin theory
with two complex scalar fields \cite{Gaberdiel:2011zw}.

\section{The minimal ${\cal N}=1$ susy extension of the bosonic cosets}\label{sec:bosex}

The above bosonic coset can be minimally extended to an ${\cal N}=1$ superconformal algebra when $k=N$, i.e.\ 
when $\lambda=\tfrac{1}{2}$. Indeed, at $k=N$ the WZW model based on $\mathfrak{su}(N)_N$ has central charge
$\tfrac{1}{2}(N^2-1)$, and can be realised in terms of $(N^2-1)$ free fermions. Thus there exists a conformal embedding
\begin{equation} 
\mathfrak{su}(N)_N \hookrightarrow \mathfrak{so}(N^2-1)_1\ , 
\end{equation}
and we can make the bosonic theory supersymmetric by considering the charge conjugation modular invariant based on
$\mathfrak{so}(N^2-1)_1$ rather than $\mathfrak{su}(N)_N$. (From the point of view of the original coset, the resulting
theory then corresponds to a `D-type modular invariant'.)

More specifically, the branching rule of the vacuum representation of $\mathfrak{so}(N^2-1)_1$ into  $\mathfrak{su}(N)_N$ 
representations is of the form
\be\label{eq3}
{\cal H}_0^{(1)} = [0^{N-1}] \  \oplus \ [2,0^{N-4},1,0] \ \oplus \ [0,1,0^{N-4},2]\  \oplus \  \cdots    \ , 
\ee
as was already shown in  \cite{Schoutens:1990xg}. Similarly, for the vector representation of $\mathfrak{so}(N^2-1)_1$
the first few terms are 
\begin{equation}\label{eq4}
{\cal H}_{v}^{(1)} = [1,0^{N-3},1] \ \oplus \ [1,1,0^{N-5},1,1] \ \oplus \ \cdots \ .
 \end{equation}
We shall momentarily explain how to describe the full decomposition series in both cases. Before we get to this we should also 
mention that  $\mathfrak{so}(N^2-1)_1$ has two spinor representations, whose conformal weight however
scales with $N$. They will therefore not play any role in the 't~Hooft limit, compare also the discussion in \cite{Gaberdiel:2011nt}.

\subsection{The branching functions}

In order to identify the full branching rules describing (\ref{eq3}) and (\ref{eq4}) we recall that the 
conformal weight of the primary in the representation $\Pi$ of $\mathfrak{su}(N)_N$ equals
\begin{equation}   \label{eq7} 
h_{\textrm{WZW}} (\Pi) = \frac{C_2(\Pi)}{2N} \ , 
\end{equation}
where the quadratic Casimir takes the form
\begin{equation}  \label{eq5}
2 \, C_2(\Pi)= \langle \Pi,\Pi+2\rho \rangle =
|\Pi|N+ \sum_{i=1}^{N-1} r_i^2 - \sum_j c_j^2-\frac{|\Pi|^2}{N} \ . 
\end{equation} 
Here $|\Pi|$ is the total number of boxes of the Young diagram corresponding to $\Pi$,
$r_j$ is the number of boxes in the $j^{\text{\tiny th}}$ row, while $c_j$ 
denotes the number of boxes in the $j^{\text{\tiny th}}$ column. In the 't~Hooft limit, it is natural to think
of $\Pi$ as being given in terms of two finite subdiagrams $\Pi_l$ and $\Pi_r$ that denote the contributions of 
the boxes and anti-boxes, respectively. In terms of these, the quadratic Casimir has the form 
\cite{Gross:1993hu}  (note that the different factor of $2$ comes from a different normalisation of the 
quadratic Casimir relative to \cite{Gross:1993hu})
\begin{equation} \label{eq6} 
C_2 (\Pi)= C_2 (\Pi_l) + C_2 (\Pi_r) + \frac{|\Pi_l| |\Pi_r|}{N}\ .
\end{equation}
This identity is actually true even at finite $N$, irrespective of how $\Pi$ is split up into $\Pi_l$ and $\Pi_r$, see 
the figure and explanation in \cite{supersymmetric}. 
The $\mathfrak{su}(N)_N$ representations that should be included in the vacuum or vector representations of $\mathfrak{so}(N^2-1)_1$ are those for which
the conformal weight in (\ref{eq7}) is an integer or a half-integer at finite $N$. (The vector representation of $\mathfrak{so}(N^2-1)_1$
has conformal dimension $\frac{1}{2}$.) Because of (\ref{eq7}) and (\ref{eq5}) it follows that
this condition is satisfied if one can represent $\Pi$ by a pair of subdiagrams $\Pi_l$ and $\Pi_r$ 
which are related to one another by transposition,
i.e.\ if $\Pi_l = \Pi_r^t$.
In the 't Hooft limit,  these are the only such representations: indeed,
representations that lead to integer or half-integer conformal weights have the property that $|\Pi_l|-|\Pi_r|$ is a multiple of $N$.
The condition that both diagrams $\Pi_l$ and $\Pi_r$ remain finite in the limit then implies  $|\Pi_l|=|\Pi_r|$.
Thus it is enough to check that the order $1$ term in (\ref{eq6}) vanishes, and it is easy to see that 
this is only the case provided that the two representations are related to one another by transposition.
In the following we shall denote this set of representations by 
\be
\Omega = \left\{ \Pi = (\Pi_l,\Pi_r) \mid \Pi_l = \Pi_r^t \right\} \ . 
\ee
Note that the first few terms in (\ref{eq3})  and (\ref{eq4}) are indeed of this form.

Next we observe that all representations in $\Omega$ are in fact in the zeroth congruence class of 
$\mathfrak{su}(N)$, i.e.\ their Dynkin labels satisfy 
\begin{equation} 
l_1+2\,  l_2 +\cdots. + (N-1) l_{N-1}  = |\Pi| \equiv |\Pi_l|-|\Pi_r| \equiv 0 \quad \textrm{mod} \, N \ .
\end{equation}
Thus the coset representations $(\Pi;0)$ with $\Pi\in\Omega$
satisfy the coset selection rule with the representation of $\mathfrak{su}(N)_1$ being 
the trivial vacuum representation; hence it is consistent to add these representations to the vacuum representation
(or to $([1,0^{N-3},1];0)$). 
Depending on whether the number of boxes $|\Pi_l|=|\Pi_r|$ is even or odd, the conformal dimension
is integer or half-integer, and hence $\Pi$  contributes to the extended vacuum  or
vector representations of $\mathfrak{so}(N^2-1)_1$, respectively, 
see (\ref{eq3})  and (\ref{eq4}). In a fermionic theory it is natural to add both of them together; the sum of 
these two representations then defines the superconformal vacuum representation. 

\noindent 
We can similarly define the extended representations 
\be\label{HLambda}
{\cal H}_\Lambda = \bigoplus_{\Pi\in\Omega} (\Pi;\Lambda) \ , 
\ee
and by construction it is clear that the conformal weights of the various different representations differ again
by integers or half-integers. On the other hand we cannot extend any of the representations of the form 
$(\Lambda_+;\Lambda_-)$  in any obvious manner, unless $\Lambda_+=0$ (or $\Lambda_+\in\Omega$). 
Thus the full extended theory is of the form
\begin{equation} \label{hilbert2} 
\mathcal{H} =\bigoplus_{\Lambda}{\cal H}_\Lambda\otimes \overline{{\cal H}}_{\Lambda^\ast}
=\bigoplus_{\Lambda}  \bigoplus_{\Pi,\Pi' \in \Omega} (\Pi; \Lambda) \otimes 
\overline{(\Pi'; \Lambda^\ast)}\ ,
\end{equation}
where the first sum runs over all pairs $\Lambda=(\Lambda_l,\Lambda_r)$ of finite Young diagrams, and 
the conjugate representation equals $\Lambda^\ast = (\Lambda_r,\Lambda_l)$.
This theory contains fermionic as well as bosonic fields (since the conformal weights
associated to the representations in $\Omega$ can be half-integer as well as integer), and the corresponding
partition function is therefore not invariant under $T:\tau\mapsto \tau+1$, but only under $T^2:\tau \mapsto \tau+2$. 
This can be cured in the usual manner by 
introducing a GSO-projection onto the bosonic states (as well as adding in the spinor representations). However, in
order to relate the theory to the higher spin dual theory, it is more natural to consider directly this fermionic theory.

\subsection{The extended characters and the spin spectrum}\label{sec:br}

Next we want to study the partition function of (\ref{hilbert2}). In particular, we want to show that the 
extended vacuum representation defines the minimal ${\cal N}=1$ superconformal extension of ${\cal W}_{\infty}[\frac{1}{2}]$.
We also need to evaluate the contributions of the other sectors in order to be able to identify them with
suitable matter field contributions on the higher spin side.

Recall from \cite{Gaberdiel:2011zw} that in the 't~Hooft limit the character of the bosonic coset representation
$ (\Pi;\Lambda)$ is of the form 
\begin{equation} \label{branching} 
b^\lambda_{\Pi;\Lambda} (q)= q^{-\frac{c}{24}} \, q^{\frac{\lambda}{2} (|\Pi_l|+|\Pi_l|-|\Lambda_l|-|\Lambda_r|)} \,
\tilde{M} (q) \sum_{\Xi} c_{\Pi \Lambda^\ast}^{\Xi} \, \textrm{ch}_{\Xi_l^t} (U_{1/2}) \cdot \textrm{ch}_{\Xi_r^t} (U_{1/2})\ , 
\end{equation}
where $\tilde{M} (q)$ is the modified MacMahon function
\begin{equation} 
\tilde{M}(q)=\prod_{n=2}^\infty \frac{1}{(1-q^n)^{n-1}}=\prod_{s=2}^\infty \prod_{n=s}^\infty \frac{1}{1-q^n}\ , 
\end{equation}
and $c_{\Pi \Lambda^\ast}^{\Xi}$ are the Clebsch-Gordan coefficients of $\mathfrak{su}(N)$. Furthermore, 
we define
\begin{equation} \label{chdef}
\textrm{ch}_\Lambda(U_h)=\sum_{T\in\textrm{Tab}_\Lambda} \prod_{j\in T} q^{h+j}\ , 
\end{equation}
where the matrix $U_h$ has only the diagonal matrix elements $(U_h)_{jj}=q^{h+j}$, and the sum is over a filling 
of the boxes of a semistandard Young tableau of shape $\Lambda$ with integers $j\geq 0$.
Thus the extended vacuum character $\chi_0$ equals 
\begin{align} 
\chi_0(q)=\sum_{\Pi\in\Omega} b^\lambda_{\Pi; 0} (q)= 
&q^{-\frac{c}{24}} \, \tilde{M}(q) \, \sum_{\Xi} q^{\frac{|\Xi|}{2}} \, 
\textrm{ch}_{\Xi}(U_{\frac{1}{2}}) \cdot \textrm{ch}_{\Xi^t}(U_{\frac{1}{2}})\nonumber\\=
& q^{-\frac{c}{24}} \tilde{M}(q) \sum_{\Xi} \textrm{ch}_{\Xi}(U_{\frac{3}{4}}) \cdot \textrm{ch}_{\Xi^t}(U_{\frac{3}{4}})\ ,
\label{eq9} \end{align}
where $\Xi$ runs over all representations with finitely many boxes. Finally, using (\ref{chdef}) as well as the  
dual Cauchy identity, see e.g.\ \cite[p.\ 65]{MacDonald}, this simplifies to 
\be \label{eq10}
\chi_0(q)=q^{-\frac{c}{24}}\, \tilde{M}(q) \, 
\prod_{r,u=0}^\infty (1+q^{r+u+\frac{3}{2}}) =
q^{-\frac{c}{24}} \, \tilde{M}(q) \, \prod_{s=1}^\infty \prod_{n=s}^\infty (1+q^{n+\frac{1}{2}})\ . 
\ee
In particular, the spin spectrum of the corresponding $\mathcal{N}=1$ ${\cal W}$-algebra consists of the bosonic currents
of spin $s=2,3,\ldots$ --- this is the contribution from the MacMahon function --- as well as fermionic currents
of spin $s=\tfrac{3}{2},\tfrac{5}{2}, \ldots$, each again with multiplicity one. Thus the total spin spectrum consists
of the minimal ${\cal N}=1$ spin content, namely of the ${\cal N}=1$ multiplets of spin 
\be\label{spincon}
(\tfrac{3}{2},2)\ ,\quad  (\tfrac{5}{2},3) \ ,  \quad (\tfrac{7}{2},4) \ , \quad \cdots  \ . 
\ee
\medskip

For general $\Lambda$ the analysis becomes a little more  involved since the Clebsch-Gordan coefficients are no longer 
trivial. If we denote by $\chi_\Lambda$ the character of the extended representation ${\cal H}_{\Lambda}$ of 
(\ref{HLambda}), we find
\begin{align}  \label{eq12} 
\chi_\Lambda(q)=&\sum_{\Pi\in\Omega} b_{\Pi;\Lambda}^\lambda(q) \nonumber\\ =
&q^{-\frac{c}{24}} \, \tilde{M}(q) \, q^{-\frac{1}{4} (|\Lambda_l|+|\Lambda_r|)}  
\sum_{\Pi\in\Omega} q^{\frac{|\Pi_l| + |\Pi_r|}{4}} \, \sum_{\Xi} c_{\Pi \, \Lambda^\ast}^{\Xi} \,
\textrm{ch}_{\Xi_l^t}(U_\frac{1}{2}) \cdot \textrm{ch}_{\Xi_r^t}(U_\frac{1}{2})\ .
\end{align}
In order to simplify these sums one can rewrite the Clebsch-Gordan coefficients of $\Pi$ and $\Lambda$ in terms of those
involving the corresponding subdiagrams $\Pi=(\Pi_l,\Pi_r)$ and $\Lambda=(\Lambda_l,\Lambda_r)$. This is discussed in 
appendix~\ref{app:B}, and it leads to (see eq.~(\ref{sub})) 
\begin{equation} \label{eq:beauty}
\chi_{(\Lambda_l,\Lambda_r)}(q) =\chi_0(q) \cdot 
\sum_{_{(\Xi_l,\Xi_r)}} c_{(\Lambda_l^t,0) (0,\Lambda_r)}^{(\Xi_l^t,\Xi_r)} \, 
\textrm{sch}_{\Xi_l^t}(\mathcal{U}_{\frac{1}{4}}) \cdot \textrm{sch}_{\Xi_r^t}(\mathcal{U}_{\frac{1}{4}})\ ,
\end{equation}
where the supercharacters are defined in eq.~\eqref{sch}.

\subsection{The full spectrum}

In order to compare to the dual higher spin theory we now need to determine  the full partition
function in the 't~Hooft limit. In the bosonic case this turned out to be somewhat subtle \cite{Gaberdiel:2011zw} 
since certain states decouple (and become null) in this limit, and therefore should not contribute to the 
partition function. We shall assume that a similar phenomenon takes place
in the present case, and that its effect amounts to replacing the Clebsch-Gordan coefficient 
\begin{equation} \label{assume}
c_{(\Lambda_l^t,0) (0,\Lambda_r)}^{(\Xi_l^t,\Xi_r)} \longrightarrow \delta^{\Xi^t_l}_{\Lambda^t_l} \, 
\delta^{\Xi_r}_{\Lambda_r}\ ,
\end{equation}
in close analogy to what happened in \cite{Gaberdiel:2011zw}.\footnote{Indeed, eq.~(\ref{assume}) simply means that 
no boxes are allowed to cancel against anti-boxes. Without this prescription, the partition function diverges in the 
't Hooft limit.} Then the character associated to $\Lambda=(\Lambda_l,\Lambda_r)$ becomes
\begin{equation} 
\chi^{\textrm{dec}}_{(\Lambda_l,\Lambda_r)}(q) = \chi_0(q) \cdot
\textrm{sch}_{\Lambda_l^t}(\mathcal{U}_{1/4}) \cdot \textrm{sch}_{\Lambda_r^t}(\mathcal{U}_{1/4})\ ,  
\end{equation}
and the full partition function equals
\begin{align} \label{part2}
\mathcal{Z}^{\textrm{dec}}_{\textrm{CFT}} = & 
\sum_{\Lambda_l,\Lambda_r} |\chi^{\textrm{dec}}_{(\Lambda_l,\Lambda_r)}|^2 \nonumber\\
= & \, |\chi_0|^2\cdot  \sum_{\Lambda_l,\Lambda_r}  \left| \textrm{sch}_{\Lambda_l}(\mathcal{U}_{1/4}) \cdot \textrm{sch}_{\Lambda_r}
(\mathcal{U}_{1/4}) \right|^2 \nonumber\\
= & \, (q\bar{q})^{-\frac{c}{24}}\, |\tilde{M}(q)|^2\, \prod_{s=1}^\infty \prod_{n=s}^\infty |1+q^{n+\frac{1}{2}}|^2 \, 
\sum_{\Lambda_l,\Lambda_r} \left| \textrm{sch}_{\Lambda_l}(\mathcal{U}_{1/4}) \cdot 
\textrm{sch}_{\Lambda_r}(\mathcal{U}_{1/4}) \right|^2\ . 
\end{align}

\section{The ${\cal N}=1$ $s{\cal W}_{\infty}$ algebra}\label{sec:sWalgebra}

Before we proceed to identify the dual higher spin theory, we first want to understand in more detail the
most general ${\cal N}=1$ superconformal $s{\cal W}_{\infty}$ algebra whose spin content agrees with 
(\ref{spincon}). As we shall see, for each value of the central charge $c$, there is a unique such algebra,
and it contains indeed the bosonic ${\cal W}_{\infty}[\tfrac{1}{2}]$ algebra as a subalgebra.  

\subsection{Structure of ${\cal N}=1$ primaries}

The analysis of the ${\cal N}=1$ superconformal $s{\cal W}$-algebra is most easily performed using ${\cal N}=1$ superfields. 
In particular, the energy-momentum tensor $T$  and the supercurrent $G$ can be combined into the superfield
$\mathbb T = \frac{1}{2}G+\theta\,T$ , whose OPE is of the form 
\be
\mathbb T(Z_{1})\mathbb T(Z_{2}) = \frac{c/6}{Z_{12}^{3}}+\frac{\frac{3}{2}\theta_{12}\mathbb T(Z_{2})}{Z_{12}^{2}}
+\frac{\frac{1}{2}D\mathbb T(Z_{2})}{Z_{12}}+\frac{\theta_{12}\mathbb T'(Z_{2})}{Z^{12}}+\cdots\ ,
\ee
where $Z_{12}=z_{1}-z_{2}-\theta_{1}\theta_{2}$, $\theta_{12}=\theta_{1}-\theta_{2}$ and 
$D=\partial_{\theta}+\theta\partial$ with $D^{2}=\partial^{2}$.
A superprimary field $\mathbb V^{(h)}=V^{(h)}+\theta\,V^{h+\frac{1}{2}}$ is similarly defined by the OPE
\be
\mathbb T(Z_{1}) \mathbb V^{(h)}(Z_{2}) = \frac{h\,\theta_{12}\mathbb V^{(h)}(Z_{2})}{Z_{12}^{2}}
+\frac{\frac{1}{2}D\mathbb V^{(h)}(Z_{2})}{Z_{12}}+\frac{\theta_{12}\,\partial\mathbb V^{(h)}(Z_{2})}{Z_{12}}+\cdots\ .
\ee
In particular, 
this implies that $V^{(h)}$ and $V^{(h+\frac{1}{2})}$ are Virasoro primaries. The super OPE between two
superprimary fields can be expanded in superconformal families that are obtained by acting with the 
negative modes of $T$ and $G$. The detailed structure can be fixed by imposing associativity with the
OPE of  $\mathbb T$; alternatively, one may use the general results 
of \cite{FigueroaO'Farrill:1990ak}.
Our conventions for the structure constants of super OPEs follow \cite{Blumenhagen:1991nm}, where useful 
selection rules have been derived.

\subsection{Enumerating superprimaries}

The other important ingredient for the analysis of the $s{\cal W}_{\infty}$ algebra is the structure of 
the various ${\cal N}=1$ superprimaries that are contained in the vacuum representation. Their numbers
can be easily determined using character techniques. To this end we expand the vacuum character of
$s{\cal W}_{\infty}$ 
in terms of ${\cal N}=1$ superconformal characters as\footnote{In this section we routinely drop the 
factor $q^{-c/24}$ from characters.}
\be
\chi_{0}(q) = \prod_{n=2}^{\infty} \frac{1+q^{n-1/2}}{1-q^n} 
+\sum_{h} d_{h}\,q^{h}
\prod_{n=1}^{\infty}\frac{1}{1-q^{n}}\,\prod_{n=1}^{\infty}(1+q^{n-\frac{1}{2}})\ ,
\ee
where the first term describes the contribution of the ${\cal N}=1$ superconformal descendants of the 
vacuum, while $d_h$ is the multiplicity of the ${\cal N}=1$  superconformal primary of conformal dimension $h$. 
Given the explicit formula for the $\mathcal N=1$ $s\mathcal W_{\infty}$ vacuum representation, see eq.~(\ref{eq10}), 
the generating function of these multiplicities turns out to equal 
\be\label{eq:dh}
\sum_{h} d_{h}\,q^{h} = q^{5/2}+q^{7/2}+q^{9/2}+2 q^{11/2}+2 q^6 +2 q^{13/2}+2 q^7+5 q^{15/2}+\cdots\ .
\ee
Apart from the algebra generators that appear at every half-integer conformal dimension with multiplicity one,
we therefore have additional composite superprimaries, the first of which has conformal dimension $\tfrac{11}{2}$. 
We shall use the convention that the former (i.e.\ the algebra generators) are denoted by 
$\mb V^{(h)}$, while the latter will be labelled as $\mb V^{(h), a}, \mb V^{(h), b}, \dots$.

\subsection{Jacobi identities}
\label{sec:jac}

With these preparations we can now discuss the 
actual construction of  $\mathcal N=1$ $s\mathcal W_{\infty}$  
by imposing recursively the Jacobi identities that encode the associativity of the operator algebra;
this can be done using the same techniques as in \cite{Candu:2012tr,Candu:2012ne}.
As will become clear, one can in principle push the analysis to arbitrary order, but obviously
the problem becomes more and more complex.
\medskip


\noindent Let us begin with the ansatz for the OPE 
\be\label{eq:55}
\super 5 \times \super 5 = \mb I+\super 7\ . 
\ee
Here we have chosen a particular normalisation for both $\super 5$ and $\super 7$.\footnote{Note 
that these choices implicitly assume that both $\super 5$ and $\super 7$ actually appear in the algebra,
as will be generically the case. However, when we discuss truncations of the 
$s\mathcal W_{\infty}$ algebra to finitely generated algebras later on, we have to be careful about choices of 
this kind.}  By dimension counting, also the superprimary $\super 9$ would have been allowed to appear in this OPE, 
but it is forbidden by the 3-point function selection rules of \cite{Blumenhagen:1991nm}.
The next OPE is
\be
\super 5 \times \super 7 = \coupling{\frac{5}{2}}{\frac{7}{2}}{\frac{5}{2}}\,\super 5+\super 9 \ ,
\ee
where the coefficient with which $\super 5$ appears is a coupling constant. In principle also
the two superprimaries with $h=\tfrac{11}{2}$ could have appeared in this OPE, but the associativity of 
$\super 5 \times \super 5 \times \super 5$ requires that they do not. In fact,
we have found empirically that the associativity constraints imply that the OPEs always 
respect the symmetry
\begin{equation}
\mathbb{V}^{(s)}\mapsto (-1)^{s+\frac{1}{2}}\mathbb{V}^{(s)} \ .
\label{eq:OPE-automorphism}
\end{equation}
This is the natural generalisation of the $W^{(s)}\mapsto (-1)^s W^{(s)}$ automorphism
symmetry of the bosonic $\mathcal{W}_\infty[\tfrac{1}{2}]$ algebra.

The associativity of $\super 5 \times \super 5 \times \super 5$ not only leads to the above selection rules, but it also 
fixes the coupling constant to equal
\be
\coupling{\frac{5}{2}}{\frac{7}{2}}{\frac{5}{2}} = \frac{192 (2 c+5) (7 c-10)}{c (4 c+21) (10 c-7)}\ .
\ee
\medskip

\noindent {\bf The OPEs $\frac{7}{2}\times \frac{7}{2}$ and  $\frac{5}{2}\times \frac{9}{2}$:}
\smallskip

\noindent Using the previous results, we can now build explicit expressions for the composite superprimaries
\be
\label{eq:first-composites}
\extrasuper{11}{a}, \qquad \mb V^{(6), a}, \qquad \mb V^{(6), b}\ . 
\ee
Schematically, they are of the form 
\begin{equation}
V^{(\frac{11}{2}), a} = (V^{(\frac{5}{2})}\,V^{(3)})+\cdots\ ,\quad 
V^{(6), a} = -\frac{1}{10}\,(V^{(3)}\,V^{(3)})+\cdots\ ,\quad
V^{(6),b} = (V^{(\frac{5}{2})}\,V^{(\frac{7}{2})})+\cdots\ ,
\end{equation}
where $(AB)$ denotes the normal ordered product, and the dots stand for terms involving $G$- and $T$-descendants.
Then we can make the most general ansatz for the next two OPEs as 
\begin{align}
\super 7 \times \super 7 &= \coupling{\frac{7}{2}}{\frac{7}{2}}{0}\,\mb I
+\coupling{\frac{7}{2}}{\frac{7}{2}}{\frac{7}{2}}\,\super 7+
\super{11}+
\coupling{\frac{7}{2}}{\frac{7}{2}}{\frac{11}{2}, a}\,\extrasuper{11}{a}+
\coupling{\frac{7}{2}}{\frac{7}{2}}{6, a}\,\extracompsuper{6}{a}+
\coupling{\frac{7}{2}}{\frac{7}{2}}{6, b}\,\extracompsuper{6}{b}\ , \nonumber \\
\super 5 \times \super 9 &=
\coupling{\frac{5}{2}}{\frac{9}{2}}{\frac{7}{2}}\,\super 7+
\coupling{\frac{5}{2}}{\frac{9}{2}}{\frac{11}{2}}\,\super{11}+
\coupling{\frac{5}{2}}{\frac{9}{2}}{\frac{11}{2}, a}\,\extrasuper{11}{a}+
\coupling{\frac{5}{2}}{\frac{9}{2}}{6, a}\,\extracompsuper{6}{a}+
\coupling{\frac{5}{2}}{\frac{9}{2}}{6, b}\,\extracompsuper{6}{b} \ .
\end{align}
Again, the superprimaries with $h=\tfrac{13}{2}$ could have appeared, but we have found that they do not,
in agreement with (\ref{eq:OPE-automorphism}). 
Imposing the associativity of $\super 5 \times \super 5 \times \super 7$ then leads to the constraints

\begin{align}
\coupling{\frac{7}{2}}{\frac{7}{2}}{0} &= \frac{192 (2 c+5) (7 c-10)}{c (4 c+21) (10 c-7)}\ , 
&\coupling{\frac{5}{2}}{\frac{9}{2}}{\frac{7}{2}} &= \frac{150 (4 c+21) (6 c-13)}{c (2 c+37) (10 c-7)}\ , \nonumber\\
\coupling{\frac{7}{2}}{\frac{7}{2}}{\frac{7}{2}} &=  \frac{720 \left(8 c^2-9 c-34\right)}{c (4 c+21) (10 c-7)}\ ,
&\coupling{\frac{5}{2}}{\frac{9}{2}}{\frac{11}{2}} &= \frac{5}{4}\ ,\nonumber \\
\coupling{\frac{7}{2}}{\frac{7}{2}}{\frac{11}{2}, a} &=  \frac{4}{5}  \,
\coupling{\frac{5}{2}}{\frac{9}{2}}{\frac{11}{2}, a}+\frac{46080 (7 c-10)}{11 c (2   c+37) (10 c-7)}\ ,
&\coupling{\frac{5}{2}}{\frac{9}{2}}{6, a} &= -\frac{7200 (14 c+11)}{11 c (2 c+37) (10 c-7)}\ ,\\
\coupling{\frac{7}{2}}{\frac{7}{2}}{6, a} &=  -\frac{11520}{11 c (10 c-7)}\ ,
&\coupling{\frac{5}{2}}{\frac{9}{2}}{6, b} &= 0\ , \nonumber\\
\coupling{\frac{7}{2}}{\frac{7}{2}}{6, b} &=  0\ .\nonumber
\end{align}
\medskip

\noindent {\bf The OPEs $\frac{7}{2}\times \frac{9}{2}$ and  $\frac{5}{2}\times\frac{11}{2}$:}
\smallskip

\noindent The ansatz for the next OPEs are 
%
\begin{align}
\super 7 \times \super 9 = &
\coupling{\frac{7}{2}}{\frac{9}{2}}{\frac{5}{2}}\,\super 5
+\coupling{\frac{7}{2}}{\frac{9}{2}}{\frac{9}{2}}\,\super 9
+\coupling{\frac{7}{2}}{\frac{9}{2}}{\frac{11}{2}}\,\super{11}
+\coupling{\frac{7}{2}}{\frac{9}{2}}{\frac{11}{2}, a}\,\extrasuper{11}{a}
+\coupling{\frac{7}{2}}{\frac{9}{2}}{6, a}\,\extracompsuper{6}{a}\nonumber \\
& + \coupling{\frac{7}{2}}{\frac{9}{2}}{6, b}\,\extracompsuper{6}{b}
+ \super{13}
+\coupling{\frac{7}{2}}{\frac{9}{2}}{\frac{13}{2}, a}\,\extrasuper{13}{a}
+\coupling{\frac{7}{2}}{\frac{9}{2}}{7, a}\,\extracompsuper{7}{a}
+\coupling{\frac{7}{2}}{\frac{9}{2}}{7, b}\,\extracompsuper{7}{b}\nonumber \\
&
+\coupling{\frac{7}{2}}{\frac{9}{2}}{\frac{15}{2}}\,\super{15}
+\coupling{\frac{7}{2}}{\frac{9}{2}}{\frac{15}{2}, a}\,\extrasuper{15}{a}
+\coupling{\frac{7}{2}}{\frac{9}{2}}{\frac{15}{2}, b}\,\extrasuper{15}{b}
+\coupling{\frac{7}{2}}{\frac{9}{2}}{\frac{15}{2}, c}\,\extrasuper{15}{c}
+\coupling{\frac{7}{2}}{\frac{9}{2}}{\frac{15}{2}, d}\,\extrasuper{15}{d}\ , \label{OPE79} \\
\super 5 \times \super{11} = &
\coupling{\frac{5}{2}}{\frac{11}{2}}{\frac{5}{2}}\,\super 5
+\coupling{\frac{5}{2}}{\frac{11}{2}}{\frac{7}{2}}\,\super 7
+\coupling{\frac{5}{2}}{\frac{11}{2}}{\frac{9}{2}}\,\super 9
+\coupling{\frac{5}{2}}{\frac{11}{2}}{\frac{11}{2}, a}\,\extrasuper{11}{a}
+\coupling{\frac{5}{2}}{\frac{11}{2}}{6, a}\,\extracompsuper{6}{a}\nonumber \\
&+\coupling{\frac{5}{2}}{\frac{11}{2}}{6, b}\,\extracompsuper{6}{b}
 +\coupling{\frac{5}{2}}{\frac{11}{2}}{\frac{13}{2}}\,\super{13}
+\coupling{\frac{5}{2}}{\frac{11}{2}}{\frac{13}{2}, a}\,\extrasuper{13}{a}
+\coupling{\frac{5}{2}}{\frac{11}{2}}{7,a}\,\extracompsuper{7}{a}
+\coupling{\frac{5}{2}}{\frac{11}{2}}{7,b}\,\extracompsuper{7}{b}\nonumber \\
&
+\coupling{\frac{5}{2}}{\frac{11}{2}}{\frac{15}{2}}\,\super{15}
+\coupling{\frac{5}{2}}{\frac{11}{2}}{\frac{15}{2},a}\,\extrasuper{15}{a}
+\coupling{\frac{5}{2}}{\frac{11}{2}}{\frac{15}{2},b}\,\extrasuper{15}{b}
+\coupling{\frac{5}{2}}{\frac{11}{2}}{\frac{15}{2},c}\,\extrasuper{15}{c}
+\coupling{\frac{5}{2}}{\frac{11}{2}}{\frac{15}{2},d}\,\extrasuper{15}{d} \label{OPE511}\ .
\end{align}
Note that we have to include superprimaries of conformal dimension $h=\tfrac{15}{2}$ since
(\ref{eq:OPE-automorphism}) only predicts that the elementary superprimary of that conformal
dimension does not arise; indeed, it turns out that one of the composite superprimaries
does indeed appear.  Imposing the associativity of 
$\super 5 \times \super 7 \times \super 7$, the above coupling constants are determined, 
see Appendix~B.1 and B.2, except for 
$\coupling{\frac{5}{2}}{\frac{9}{2}}{\frac{11}{2},a}$
and $\coupling{\frac{5}{2}}{\frac{11}{2}}{\frac{13}{2},a}$. However, these two couplings are not actually 
free parameters, but just reflect the freedom that we can redefine superprimaries of the same dimension,
see \cite{Candu:2012ne} for a similar phenomenon. For the following we shall choose the convention that 
\be\label{simpl}
\coupling{\frac{5}{9}}{\frac{9}{2}}{\frac{11}{2},a} = 
\coupling{\frac{5}{9}}{\frac{11}{2}}{\frac{13}{2},a} = 0\ ,
\ee
where  the top component of $\mathbb V^{(\frac{13}{2}),a}$ is of the form
\begin{equation}
\label{eq:second-composite}
V^{(\frac{13}{2}),a} = (V^{(\frac{5}{2})}\,V^{(4)})+\frac{7}{5}(V^{(3)}V^{(\frac{7}{2})})+\cdots
\ .
\end{equation}
We have checked that the associativity of $\super 5 \times \super 5 \times \super 9$ is then also
satisfied by these OPEs.
\smallskip

\noindent We have also analysed the Jacobi identities involving the OPEs of 
\begin{equation}
\super 7 \times \super{11} \ ,  \qquad \super 9 \times \super 9 \ , \qquad
\super 5 \times \super{13} \ , 
\end{equation}
and determined the relevant structure constants. Some of the details of this analysis are given
in Appendices~B.3 and B.4. 
\smallskip

In summary, these considerations therefore suggest that the algebra does not have 
any free parameter beyond the central charge. In the following three subsections we shall
subject these results to some independent consistency checks. First, in Section~\ref{sec:trunc},
we shall show that the algebra truncates to the finitely generated ${\cal N}=1$ algebras
for the appropriate values of the central charge. In the remaining two subsections we shall
then demonstrate that it contains indeed the bosonic ${\cal W}_{\infty}[\tfrac{1}{2}]$ algebra
as a subalgebra, first directly in terms of the structure constants, and then by studying the
so-called minimal representations that will also play a role for the higher spin holography.

\subsection{Truncation properties}\label{sec:trunc}

We have seen in Section~\ref{sec:br} that the vacuum character of the coset reproduces the spectrum of 
$s\mathcal{W}_\infty$ in the $N\to\infty$ limit.
It is therefore natural to expect that when the central charge  equals one of the coset values
\begin{equation}
c_N = \frac{(3N+1)(N-1)}{2(2N+1)}
\label{eq:cncoset}
\end{equation}
the $s\mathcal{W}_\infty$ algebra truncates to the $\mathcal{N}=1$ extended coset algebra, which we 
denote by $s\mathcal{W}_N$. This is an exceptional $\mathcal{W}$-algebra with spin content\footnote{Let us 
mention that the direct construction of these algebras is complicated by the fact that the Jacobi identities 
of the generators can only be satisfied modulo non-trivial null fields.}
$$
(\tfrac{3}{2},2)\ ,\quad (\tfrac{5}{2},3)\ ,\quad \dots\ ,\quad  (N-\tfrac{1}{2},N)\ ,
$$
see \cite{Schoutens:1990xg}.
With the explicit values of the structure constants of $s\mathcal{W}_\infty$ at hand, let us check whether the latter 
truncates to $s\mathcal{W}_N$ at $c=c_N$. In order for this to happen, it is necessary that 
\begin{equation}
\coupling{s}{s_1}{s_2} = 0 \ , \qquad \forall \ s_1\geq N+\tfrac{1}{2} \ , \quad s_2\leq N-\tfrac{1}{2} \ ,
\label{eq:trunc1}
\end{equation}
irrespective of the value of $s$. 
Similarly, for any composite field $\mathbb{V}^{(s_2),a}$ which is not part of the ideal, we must require that 
\begin{equation}
\coupling{s}{s_1}{s_2,a} = 0 \ ,
\label{eq:trunc2}
\end{equation}
provided that $s_1\geq N+\tfrac{1}{2}$. Let us now consider individually the cases 
$N=3,4,5,6$.\footnote{The case $N=2$ is a bit unnatural since the $\mathcal{N}=1$ Virasoro algebra is always
a consistent subalgebra of $s\mathcal{W}_\infty$, irrespective of any such truncation. However, if one formally 
applies the above conditions for $N=2$ one finds the two values $c=0$ and $c=c_2=\tfrac{7}{10}$.}
%
%
Demanding (\ref{eq:trunc1}) for $N=3$ requires in particular that 
\begin{equation}
\coupling{\frac{7}{2}}{\frac{7}{2}}{0}=\coupling{\frac{9}{2}}{\frac{9}{2}}{0}
=\coupling{\frac{5}{2}}{\frac{7}{2}}{\frac{5}{2}}=\coupling{\frac{5}{2}}{\frac{11}{2}}{\frac{5}{2}}=0 \ ,
\end{equation}
which leads to $c=-\tfrac{5}{2}$ or $c=c_3 = \tfrac{10}{7}$, in perfect agreement with \cite{Inami:1988xy}.
For the $N=4$ truncation we need to set
$$
\coupling{\frac{9}{2}}{\frac{9}{2}}{0}=\coupling{\frac{5}{2}}{\frac{9}{2}}{\frac{7}{2}}=\coupling{\frac{5}{2}}{\frac{11}{2}}{\frac{5}{2}}=
\coupling{\frac{7}{2}}{\frac{9}{2}}{\frac{5}{2}}=\coupling{\frac{7}{2}}{\frac{11}{2}}{\frac{7}{2}}=\coupling{\frac{5}{2}}{\frac{9}{2}}{\frac{11}{2},a}=\coupling{\frac{5}{2}}{\frac{11}{2}}{\frac{13}{2},a}=0\ ,
$$
which gives, using (\ref{eq:7117}) and (\ref{eq:990}),
$c=c_4=\tfrac{13}{6}$. In order to deal with $N\geq 5$ we need to relax 
eq.~\eqref{simpl} and fix the values of the respective structure constants from the truncation analysis.
We have found that a necessary condition for the truncation to $N=5$ to occur is that $c$ equals either
\begin{equation}
c = - \frac{2}{5} \ , \quad \hbox{or} \quad c = - \frac{16}{5} \ , \quad \hbox{or}  \quad
c = \frac{32}{11} \equiv c_5 \ . 
\end{equation}
Finally, we have checked that for $N=6$, $c=c_6$ is a solution of the truncation constraints (but
we have not determined sufficiently many structure constants in order to rule out a number of other solutions).

\subsection{Bosonic subalgebra}\label{sec:bossub}

The $\mathcal{N}=1$ extension of the cosets~\eqref{bosoniccoset} with $k=N$ implies that the chiral algebra of the bosonic coset must be  a subalgebra of its extension, i.e.\ $\mathcal{W}_N\subset s\mathcal{W}_N$.
In this section we give strong evidence for the claim that the same subalgebra structure lifts to the infinitely generated algebras
\begin{equation}
\mathcal{W}_\infty[\tfrac{1}{2}]\subset s\mathcal{W}_\infty\  .
\end{equation}

To set up the notation, recall that the $\mathcal W_{\infty}[\mu]$ algebra is generated by primary fields
$W^{(s)}$ with dimension $s=3, 4, 5, \dots$, that we normalise as
$W^{(s)}\times W^{(s)} = \frac{c}{s}\,\mathbb I+\cdots$.
The  first few OPEs of $\mathcal W_{\infty}[\mu]$ are 
\begin{align}
W^{(3)}\times W^{(3)} &= \frac{c}{3}\,\mathbb I + C_{33}^{4}\,W^{(4)}\ , \label{334} \\
W^{(3)}\times W^{(4)} &= C_{34}^{3}\,W^{(3)}+C_{34}^{5}\,W^{(5)}\ , 
\end{align}
where (see \cite{Hornfeck:1992tm,Gaberdiel:2012ku})
\begin{align}
(C_{33}^{4})^{2} &= \frac{64\,(c+2)(\mu-3)(c(\mu+3)+2(4\mu+3)(\mu-1))}
{(5c+22)(\mu-2)(c(\mu+2)+(3\mu+2)(\mu-1))}\ , \\
\label{eq:embedding0}
C_{34}^{3} &= \tfrac{3}{4} C_{44}^{3}\ , \\
C_{33}^{4}\,C_{44}^{4} &= \frac{48(c^{2}(\mu^{2}-19)+3c(6\mu^{3}-25\mu^{2}+15)+2(\mu-1)(6\mu^{2}-41\mu-41))}{
(\mu-2)(5c+22)(c(\mu+2)+(3\mu+2)(\mu-1))}\ , \\
(C_{34}^{5})^{2} &= \frac{25(5c+22)(\mu-4)(c(\mu+4)+3(5\mu+4)(\mu-1))}
{(7c+114)(\mu-2)(c(\mu+2)+(3\mu+2)(\mu-1))}\ .
\end{align}
These expressions reduce, for $\mu=\frac{1}{2}$, to 
\begin{align}
\label{eq:embedding1}
(C_{33}^{4})^{2} &=\frac{640 (c+2) (7 c-10)}{3 (5 c+22) (10 c-7)}\ ,  \\
\label{eq:embedding2}
C_{33}^{4}\,C_{44}^{4} &= \frac{96 \left(25 c^2-38 c-80\right)}{(5 c+22) (10 c-7)}\ , \\
\label{eq:embedding3}
(C_{34}^{5})^{2} &= \frac{175 (5 c+22) (6 c-13)}{(7 c+114) (10 c-7)}\ .
\end{align}
Since the $s{\cal W}_{\infty}$ algebra contains only a single Virasoro primary field of conformal dimension $3$, 
we must have the identification 
\be
W^{(3)} = \sqrt\frac{c}{3}\,V^{(3)}\ ,
\ee
where the prefactor is fixed by the normalisation conventions.
Evaluating the $W^{(3)}W^{(3)}$ OPE and decomposing it in terms of conformal families of primary fields, we find 
\be
W^{(3)}\times W^{(3)} = \frac{c}{3}\mathbb I + \wt W^{(4)}\ ,
\ee
where $\wt W^{(4)}$  is the Virasoro primary field 
\begin{align}
\wt W^{(4)} = & \frac{8 (7 c-10) (G'G)}{(4 c+21) (10 c-7)}-\frac{136 (7 c-10) (TT)}{(4 c+21) (5 c+22) (10 c-7)}\nonumber \\[2pt]
&-\frac{12   (c+1) (7 c-10) T''}{(4 c+21) (5 c+22) (10 c-7)}+\frac{c V^{(4)}}{3}\ .
\end{align}
Given the normalisation conventions of ${\cal W}_{\infty}[\tfrac{1}{2}]$, the correctly normalised spin $4$ field is then
\be
W^{(4)} = \sqrt\frac{3(5c+22)(10c-7)}{640(c+2)(7c-10)}\, \wt W^{(4)}\ ,
\ee
in agreement with (\ref{eq:embedding1}). Next, we can evaluate the OPE $W^{(3)}W^{(4)}$ and find that it 
takes the form 
\be
W^{(3)}\times W^{(4)} = C_{34}^{3}\,W^{(3)} + \wt W^{(5)}\ ,
\ee
where the value of $C_{34}^{3}$ is indeed in agreement with (\ref{eq:embedding0}). Note that no Virasoro
primary field of spin $s=6$ appears in this OPE, again as expected from ${\cal W}_{\infty}[\tfrac{1}{2}]$. The field
$\wt W^{(5)}$ is explicitly given as 
\begin{align}
\wt W^{(5)} = & \frac{\sqrt\frac{c(5c+22)(10c-7)}{10(c+2)(7c-10)}}{24(2c+37)(7c+114)(10c-7)}\bigg[
-360(6c-13)(7c+114) (G {V^{(\frac{5}{2})}}')\nonumber \\
&+600(6c-13)(7c+114)(G' V^{(\frac{5}{2})})
+360(c+34)(6c-13) {V^{(3)}}''\nonumber \\
&-29760(6c-13)(T V^{(3)}) +c(2c+37)(7c+114)(10c-7) V^{(5)}\bigg]\ ,
\end{align}
where $V^{(5)}$ is the bosonic component of $\super 9$. Again, all the OPEs required to normalise this field
are available, and we find
\be
W^{(5)} = \sqrt\frac{(7c+114)(10c-7)}{175(5c+22)(6c-13)}\wt W^{(5)}\ ,
\ee
in agreement with (\ref{eq:embedding3}). Finally, we can evaluate $C_{44}^{4}$ in the $W^{(4)}W^{(4)}$ OPE
\be
W^{(4)}\times W^{(4)} = \frac{c}{4}\mathbb I+C_{44}^{4} W^{(4)}+\cdots \ ,
\ee
and compute 
\be
C_{44}^{4} = \frac{6 \sqrt{\frac{6}{5}} (c (25 c-38)-80)}{\sqrt{(c+2) (5 c+22) (7 c-10) (10 c-7)}}\ ,
\ee
again in agreement with (\ref{eq:embedding2}). These checks therefore provide very convincing evidence that
$s{\cal W}_{\infty}$ contains indeed ${\cal W}_{\infty}[\tfrac{1}{2}]$ as a subalgebra.

\subsection{Minimal representations}

As an independent consistency check of the analysis of the previous subsection, we can also determine
the structure of the so-called minimal representations. They are characterised by the property that their
character equals
\begin{equation}
\chi_{\text{min}}(q)=q^{h}\, \frac{1+q^{\frac{1}{2}}}{1-q} \cdot \chi_{0} (q)\ .
\label{eq:minch}
\end{equation}
In particular, this means that the representation is highly degenerate, e.g.\ the only descendant at conformal
dimension $h+\tfrac{1}{2}$ is the $G_{-1/2}$-descendant, the only descendant at conformal dimension $h+1$
the $L_{-1}$-descendant, etc. The minimal representations play an important role in the holographic duality since
they correspond to the smallest matter multiplet.

In terms of OPEs, the representation generated from the super-Virasoro primary $\mathbb{P}^{(h)}$ defines
a minimal representation provided that its OPEs are of the form 
\be\label{eq:opemin}
\mathbb V^{(s)}\times \mathbb P^{(h)} = \sum_{\mathbb P'\in \mathscr P} 
\coupling{s}{h}{\mathbb P'} \,\,\mathbb P'  \ , 
\ee
where the superprimaries appearing in $\mathscr P$ are built out of normal ordered products of 
$\mathbb{V}^{(s)}$, $\mathbb{P}^{(h)}$ and their derivatives, that are \emph{linear} in $\mathbb P^{(h)}$ 
(including $\mathbb P^{(h)}$ itself). Indeed, this is equivalent to requiring  that all the fields 
$\{V^{(s)}_{-m}P^{(h)}\mid m<s\}$ that appear in the singular part of the 
OPE~\eqref{eq:opemin} can be expressed in 
terms of $L_{-1}$ and $G_{-1/2}$ descendants of $P^{(h)}$, or of states that 
are obtained from 
$P^{(h)}$ by the action of non-wedge modes,  $V^{(t)}_{-n}$ with $n\geq t$; this in turn
is equivalent to the characterisation of the minimal representations given in (\ref{eq:minch}). 

To determine the conformal weight of the minimal representation we can proceed as in  \cite{Candu:2012tr,Candu:2012ne}:
we make the most general ansatz of the form (\ref{eq:opemin}), and check that it is associative with respect to further 
OPEs with $\mathbb V^{(s')}$.  In order
to determine which fields may appear in (\ref{eq:opemin}), we count the super-Virasoro primaries that appear 
in the representation generated from $\mathbb{P}^{(h)}$; using similar arguments as around eq.~(\ref{eq:dh}), they are 
counted by the generating function
\begin{equation}
\sum_{h'} \hat{d}_{h'}\,q^{h'} 
= q^{h}\,\bigg[1+q^{5/2}+q^3+2 q^{7/2}+2 q^4+3 q^{9/2}+3 q^5+\cdots  \bigg]\ .
\end{equation}
The first few OPEs are rather simple. For example, for $s=\tfrac{5}{2}$ we have simply
\be
\super 5\times \mathbb P^{(h)} = \coupling{\frac{5}{2}}{h}{h}\,\mathbb P^{(h)} \ , 
\ee
since no composite superprimary can occur. For $s=\tfrac{7}{2}$ we find on the other hand 
\be
\super 7 \times \mathbb P^{(h)} = \coupling{\frac{7}{2}}{h}{h}\,\mathbb P^{(h)}+
 \coupling{\frac{7}{2}}{h}{h+\frac{5}{2}}\,\mathbb P^{(h+\frac{5}{2})}+
  \coupling{\frac{7}{2}}{h}{h+3}\,\mathbb P^{(h+3)} \ ,
\ee
where $\mathbb P^{(h+\frac{5}{2})}$  and $\mathbb P^{(h+3)}$ are the composite superprimaries of conformal
dimension $h+\tfrac{5}{2}$ and $h+3$, respectively. (It is not hard to write down explicit expressions for these composite
fields,  but we shall refrain from doing so here.) Then we can impose the associativity 
$\super{5}\times \super{5}\times \mathbb P^{(h)}$, from which it follows that the conformal dimension of $\mathbb P^{(h)}$ 
must satisfy
\be
\label{eq:ground-state-dim}
2h\,(h+2)+c\,(4h-1)=0 \ .
\ee
Furthermore, the leading couplings (from which one can read of the eigenvalues of the respective
zero modes) equal
\begin{align}
\bigg(\coupling{\frac{5}{2}}{h}{h}\bigg)^{2} &= -\frac{(h+1)^2 (4 h-1) (10 h-1)}{h (h+2) (2 h+7)}\ , \\[4pt]
\coupling{\frac{7}{2}}{h}{h} &= -\frac{12 (2 h-5) (2 h-1) (2 h+3) (4 h-1) (7 h-1)}{h (h+2) (2 h+7) \left(8 h^2-68 h+21\right)}\ .
\end{align}
The equation for the conformal dimension \eqref{eq:ground-state-dim} has two solutions, namely 
\begin{equation}
h_\pm  = - (1+c) \pm \frac{1}{2} \sqrt{(c+2)(c+\tfrac{1}{2})} \ ,
\end{equation}
which agree precisely with the two solutions given in eq.~(3.8) of \cite{Gaberdiel:2012ku} for $\mu=\tfrac{1}{2}$. 
For $c=c_N$, see eq.~\eqref{eq:cncoset}, they simplify to 
\begin{equation}
h_+ = \frac{N-1}{2(2N+1)} \equiv h(0;\mathrm{f}) \ , \qquad \hbox{and} \qquad 
h_{-} = -\frac{3N+1}{2} \ . 
\end{equation}
The first solution agrees therefore with the $(0;{\rm f})$ solution of the bosonic ${\cal W}_{\infty}$ algebra
at $\mu=\tfrac{1}{2}$. This ties in nicely with the fact that it follows from eq.~(\ref{eq:beauty}) that\footnote{Note
the potentially confusing notation $(0;{\rm f}) \cong ({\rm f},0)$, where the latter refers to the
extended representation, see (\ref{HLambda}), with $\Lambda_l={\rm f}$ and $\Lambda_r=0$.}
\begin{equation} 
\chi_{({\rm f},0)} (q) = \chi_0(q) \cdot \textrm{sch}_{\Yboxdim4pt \yng(1)}(\mathcal{U}_{1/4}) 
= \chi_0(q) \cdot \frac{q^{\frac{1}{4}}(1+q^{\frac{1}{2}})}{(1-q)} 
\end{equation}
defines indeed a minimal representation in the 't~Hooft limit. (Note that in the 't~Hooft limit, 
$h_+=\tfrac{1}{4}$.) These results provide an independent check of our claim that 
$\mathcal{W}_\infty[\tfrac{1}{2}]$ is a subalgebra of  $s\mathcal{W}_\infty$ (for all
values of the central charge).

\section{The dual higher spin point of view}\label{sec:AdShs}

We now finally turn to the description of the dual higher spin gravity theory. We begin by describing
the underlying higher spin algebra.

\subsection{The higher spin algebra}

A higher spin algebra corresponding to the spectrum (\ref{spincon}) was constructed some time ago in \cite{fradkin}.
This algebra may be described as a certain 
restriction of the ${\cal N}=2$ shs$[\mu]$ algebra at the special point $\mu=\tfrac{1}{2}$. Recall that the 
$\mathcal{N}=2$ supersymmetric higher spin algebra shs$[\mu]$ is a one parameter family of Lie superalgebras 
\cite{structure} which can be defined as the quotient
\begin{equation}  \label{casimir} 
\textrm{shs}[\mu] \oplus \mathds{C} = \frac{U(\mathfrak{osp}(1|2))}{\langle 
C^{\mathfrak{osp}(1|2)}-\frac{1}{4} \mu (\mu-1)\bold{1} \rangle}
\ ,
\end{equation}
where $U(\mathfrak{osp}(1|2))$ denotes the universal enveloping algebra  of 
$\mathfrak{osp}(1|2)$, and $C^{\mathfrak{osp}}$ is the corresponding quadratic Casimir. In order
to describe this more explicitly, let us define the  oscillator algebra 
\begin{equation} \label{ycomm}
{} [y_\alpha,y_\beta]=2i \epsilon_{\alpha \beta} (1+\nu k)\ , \qquad \{k,y_\alpha \}=0\ , \qquad k^2=\bold{1}\ ,
 \end{equation} 
where $\alpha,\beta=1,2$, 
$\frac{\nu}{2}=\left(\mu -\frac{1}{2} \right)$, and $\epsilon_{12}=+1=-\epsilon_{21}$.
Then  the quotient~\eqref{casimir} can be realised by identifying the generators of $\mathfrak{osp}(1|2)$ with
\begin{align*}
G_{\frac{1}{2}} &= \frac{1}{2}e^{-i\frac{\pi}{4}}y_1\ ,& G_{-\frac{1}{2}} &= \frac{1}{2}e^{-i\frac{\pi}{4}}y_2\ ,& 
L_1&= \frac{i}{2}y_1^2\ ,& L_{-1}&= \frac{i}{2}y_2^2\ , & L_0 &=\frac{i}{4}(y_1y_2+y_2y_1) \ , 
\end{align*}
since
\begin{equation}
C^{\mathfrak{osp}(1|2)} = L_0^2 - \frac{1}{2}\{L_1,L_{-1}\}+\frac{1}{4}[G_{\frac{1}{2}},G_{-\frac{1}{2}}]
=\frac{\nu^2}{4}\, k^2 - \frac{1}{16}\ .
\end{equation}
The elements generating shs$[\mu]$ can thus be written as symmetric products of the oscillators 
$y_{\alpha_i}$, $\alpha_i=1,2$ \cite{structure}
\begin{equation} \label{shs}
V_m^{(s)\pm} = y_{(\alpha_1...}\, y_{\alpha_n)}(\bold{1} \pm k)\ , 
\end{equation}
where $V_m^{(s)}$ has `spin' $s=1+\frac{n}{2}$ with $n\geq 0$ --- for $n=0$, $V_m^{(1)\pm}\equiv \pm k$, since
the ${\bf 1}$ generator is central and is not part of shs$[\mu]$, see (\ref{casimir}). 
Here $m$ takes the values  $2m=N_1-N_2$, where $N_{1,2}$ is the number of $y_{1,2}$, and hence lies in the range
$-s+1\leq m \leq s-1$. 
For $\mu=\tfrac{1}{2}$, i.e.\ $\nu=0$, it is consistent \cite{structure} to
restrict the generators of shs$[\mu]$ to the $k$ independent part of 
eq.~(\ref{shs}) since $k$ is never generated by any commutators, see eq.~(\ref{ycomm}). 
This construction obviously leads to an algebra that is generated by
\be\label{eq:prop}
V^{(s)}_m \propto V^{(s)+}_{m} + V^{(s)-}_{m} \ ,
\ee
i.e.\ each spin $s\ge \frac{3}{2}$ appears only once. This resulting algebra, which we shall denote
as shs$(1|2)$ as in  \cite{Blencowe:1988gj},  is isomorphic to the 
symplecton higher spin algebra shs$'_{\rho}(1)$ of \cite{fradkin}, as was pointed out in \cite{structure}. In 
\cite{fradkin}  the strategy for the construction of this algebra was different since they used a more geometric approach;
the algebra was then subsequently employed in \cite{Blencowe:1988gj} to construct a consistent action in $d$=2+1,
using a Chern-Simons action based on shs$(1|2) \oplus$ shs$(1|2)$. 

For the following it will also be important to understand the simplest matter field of this theory.
Because $\mathfrak{osp}(1|2)$ is a subalgebra of shs$(1|2)$ and shs$(1|2)$ is a quotient of $U(\mathfrak{osp}(1|2))$,
any irreducible representation
of $\mathfrak{osp}(1|2)$ for which the quadratic Casimir takes the value $C^{\mathfrak{osp}(1|2)} =-\tfrac{1}{16}$, 
see eq.~(\ref{casimir}), 
leads to an irreducible representation of shs$(1|2)$.  On a highest weight state $|h\rangle$ of  $\mathfrak{osp}(1|2)$ 
the quadratic Casimir equals 
\begin{equation} 
C^{\mathfrak{osp}(1|2)} |h\rangle =h \left(h-\frac{1}{2} \right) |h\rangle\ , 
\end{equation}
and hence $h=\tfrac{1}{4}$ leads to $C^{\mathfrak{osp}(1|2)} =-\tfrac{1}{16}$. The character of this `minimal'
representation is of the form 
\begin{equation}
\chi_{\rm min}^{\rm shs}(q) = \frac{q^{\frac{1}{4}}(1+q^{\frac{1}{2}})}{(1-q)} \ ,
\end{equation}
i.e.\ it agrees with the `wedge part' of the minimal representation (\ref{eq:minch}).


\subsection{The higher spin algebra as the wedge algebra}


Based on the general philosophy of \cite{Gaberdiel:2011wb}, one should expect that the Lie superalgebra shs$(1|2)$ agrees
precisely with the so-called wedge subalgebra of $s{\cal W}_{\infty}$; the latter is generated by the wedge modes 
$\{V^{(s)}_m\mid |m|<s\}$  in the $c\to\infty$ limit.
In the following we want to confirm that this expectation is indeed borne out.

Turning the OPEs of Section~\ref{sec:jac} into (anti-)commutators following \cite{Bowcock:1990ku}, 
restricting to the wedge, and finally taking the $c\to\infty$ limit, we find
\begin{equation}\label{eq:comWedge}
[V^{(s)}_m,V^{(s')}_{m'}] = \sum_{s''} P^{ss'}_{s''}(m,m')\, d^{ss'}_{s''} \, V^{(s'')}_{m+m'}\ ,
\end{equation}
where we have set $V^{(\frac{3}{2})}\equiv \frac{1}{2}\, G$ and $V^{(2)}\equiv T$, and it is understood that
the left-hand-side is a commutator or anti-commutator as appropriate.
The polynomials $P^{ss'}_{s''}(m,m')$ only describe the mode dependence of the commutators
and are entirely fixed by global conformal symmetry
\begin{equation}\label{eq:uni_pol}
P^{j+1\ j'+1}_{j''+1}(m,m') =\!\!\sum_{r=0}^{j+j'-j''}\binom{j+m}{j+j'-j''-r}\frac{ (-1)^r (j-j'+j''+1)_{(r)}(j''+m+m'+1)_{(r)} }{r! (2j''+2)_{(r)}}
\end{equation}
where $(a)_{(n)}=\Gamma(a+n)/\Gamma(a)$ is the Pochhammer symbol.
All the  non-trivial information about the algebra is contained in the structure constants
\begin{equation}
 d^{ss'}_{s''} =\lim_{c\to\infty}D^{ss'}_{s''}\ ,
\label{eq:corr1}
\end{equation}
where $D^{ss'}_{s''}$ are the structure constants  of the different component fields; in terms of the 
superprimary constants we have the relations, see \cite{Blumenhagen:1991nm}
\begin{equation}
\mathbb{D}^{ss'}_{s''}=2 s'' D^{s\,s'}_{s''+\frac{1}{2}} = D^{s+\frac{1}{2}\,s'}_{s''}
= (-1)^{2s+1}D^{s\,s'+\frac{1}{2}}_{s''}= \frac{(-1)^{2s+1} 2s''}{s+s'+s''-\frac{1}{2}}D^{s+\frac{1}{2}\,s'+\frac{1}{2}}_{s''+\frac{1}{2}}\ .
\end{equation}
Furthermore, the coupling to the energy momentum tensor multiplet $\mathbb{V}^{\frac{3}{2}}$ is determined by    
\begin{equation}
\mathbb{D}^{ss'}_{\frac{3}{2}} = \delta_{ss'} \frac{6s}{c} \mathbb{D}^{ss}_0\ .
\end{equation}
To the extent to which we have determined the commutation relations of $s{\cal W}_{\infty}$, we have verified that the
commutation relations~\eqref{eq:comWedge} reproduce those of  the Lie superalgebra shs$(1|2)$ provided 
the $\mathfrak{osp}(1|2)$ highest weight components are identified as 
\begin{equation}
V{}^{\frac{5}{2}}_{\frac{3}{2}} = -\frac{e^{\frac{3\,\pi\,i}{4}}}{4\,\sqrt{10}}\,y_1^3\ ,\qquad
V{}^{\frac{7}{2}}_{\frac{5}{2}} = \frac{e^{\frac{5\,\pi\,i}{4}}}{10}\,y_1^5\ ,\qquad
V{}^{\frac{9}{2}}_{\frac{7}{2}} = -\frac{e^{\frac{7\,\pi\,i}{4}}}{2\,\sqrt{10}}\,y_1^7\ .
\end{equation}
This provides very convincing evidence for the fact that the wedge subalgebra of $s{\cal W}_{\infty}$ 
is indeed isomorphic to shs$(1|2)$.

\subsection{The thermal partition function of the dual higher spin theory}\label{sec:part}

Finally, we want to compare the thermal partition function of the dual higher spin theory to the 
't~Hooft limit of the CFT partition function. First recall that the contribution of a bosonic higher spin field
of spin $s$ to the thermal 1-loop partition function on AdS equals~\cite{Gaberdiel:2010ar}
\begin{equation} 
Z_B^{(s)}=\prod_{n=s}^\infty \frac{1}{|1-q^n|^2}\ , 
\end{equation}
while that of a half-integer  spin $s$ fermion is \cite{Creutzig:2011fe}
\begin{equation} 
Z_F^{(s)}=\prod_{n=s-\frac{1}{2}}^\infty |1+q^{n+\frac{1}{2}}|^2\ . 
\end{equation}
The vacuum character of eq.~(\ref{part2}) is then precisely reproduced by a tower of massless bosonic and 
fermionic gauge fields 
\begin{equation} 
\mathcal{Z}_{\textrm{gauge}}= \prod_{s=2}^\infty Z_B^{(s)}  Z_F^{(s-\frac{1}{2})}\ , 
\end{equation}
with a spin content of $\tfrac{3}{2}, 2, \tfrac{5}{2}, 3, \tfrac{7}{2},\ldots$. This is as expected since the currents of the
conformal field theory should correspond to the massless gauge fields of the higher spin theory.

In order to account for the additional contributions to the partition function, we now need to add matter fields
to the higher spin theory. It was shown in \cite{Giombi:2008vd} that a massive complex scalar field contributes 
\begin{equation} 
Z_{\textrm{c-scalar}}^h=\prod_{j,j'=0}^\infty \frac{1}{(1-q^{h+j}\bar{q}^{h+j'})^2}\ , 
\end{equation}
where the mass equals $M^2+1=(\Delta-1)^2=(2h-1)^2$ in terms of the conformal dimension $h$ of the 
boundary excitation. Similarly,
a massive Dirac fermion whose dual conformal dimension equals $(h+\frac{1}{2},h)$ and $(h,h+\frac{1}{2})$ leads to 
\cite{Creutzig:2011fe} 
\begin{equation} 
Z_{\textrm{spinor}}^h=\prod_{j,j'=0}^\infty (1+q^{h+j}\bar{q}^{h+\frac{1}{2}+j'}) (1+q^{h+\frac{1}{2}+j}\bar{q}^{h+j'}) \ , 
\end{equation}
where the mass squared is given by $m^2=(\Delta_F-1)^2=\left(2h-\frac{1}{2} \right)^2$ with 
$\Delta_F=2h+\frac{1}{2}$ the total scaling dimenison.

\noindent Next we observe, using the same techniques as in \cite{supersymmetric}, that 
\begin{align} 
\frac{1}{\textrm{sdet}{(1-\mathcal{U}_h \otimes \mathcal{U}^*_h)}}
&=\sum_{\Xi} \textrm{sch}_\Xi(\mathcal{U}_h) \cdot \overline{\textrm{sch}_\Xi(\mathcal{U}_h)} \nonumber\\
&= \prod_{j,j'=0}^\infty \frac{(1+q^{h+j}\bar{q}^{h+\frac{1}{2}+j'}) 
(1+q^{h+\frac{1}{2}+j}\bar{q}^{h+j'})}{(1-q^{h+j}\bar{q}^{h+j'}) (1-q^{h+\frac{1}{2}+j}\bar{q}^{h+\frac{1}{2}+j'})}\ , \label{sdet} 
\end{align}
where sdet denotes the superdeterminant. Thus, up to the contribution from the gauge fields ${\cal Z}_{\rm gauge}$, 
the whole conformal field theory partition function,
eq.~(\ref{part2}), equals exactly  the square of eq.~(\ref{sdet}) with $h=\frac{1}{4}$. In terms of 
matter fields, on the other hand, the square of eq.~(\ref{sdet}) just describes the ${\cal N}=1$ 
matter multiplet consisting of   two complex scalars of mass squared $M^2=-\frac{3}{4}$, 
one with conformal dimension $(\frac{1}{4},\frac{1}{4})$, and one with $(\frac{3}{4},\frac{3}{4})$, as well as two
massless Dirac fermions each of conformal dimension 
$(\frac{3}{4},\frac{1}{4})$ and $(\frac{1}{4},\frac{3}{4})$. Thus the higher spin theory consists of the 
higher spin gauge fields of spin $s=\tfrac{3}{2},2,\tfrac{5}{2},3,\ldots$, together with the ${\cal N}=1$
matter multiplet 
\begin{equation} 
\mathcal{Z}^h_{\textrm{matter}}= Z_{\textrm{c-scalar}}^{h}(Z_{\textrm{spinor}}^h)^2 Z_{\textrm{c-scalar}}^{h+\frac{1}{2}}
\end{equation}
with $h=\tfrac{1}{4}$.

\section{Conclusions}\label{sec:concl}

In this paper we have studied the minimal ${\cal N}=1$ superconformal $s{\cal W}_{\infty}$ theory, and identified its 
higher spin dual. In particular, we have analysed the structure of the most general $s{\cal W}_{\infty}$ algebra
that is generated by one field of each half-integer spin $s\geq \tfrac{3}{2}$, and we have found that 
it does not have any free parameter, except for the central charge. We have also shown that $s{\cal W}_{\infty}$ contains 
the bosonic ${\cal W}_\infty[\tfrac{1}{2}]$ algebra as a subalgebra, and indeed the ${\cal N}=1$ theory can be obtained
by extending the bosonic ${\cal W}_\infty[\mu]$ theory at $\mu=\tfrac{1}{2}$. From that point of view, the ${\cal N}=1$
theory is described by a non-diagonal modular invariant of the bosonic ${\cal W}_{\infty}[\tfrac{1}{2}]$ algebra. 

The corresponding higher spin theory can be described in terms of a Chern-Simons theory based on the algebra
${\rm shs}(1|2)$. (Note that the ${\cal N}=1$ higher spin theory considered in \cite{Creutzig:2012ar} has a 
different spin content and is instead described by a truncation of the ${\cal N}=2$ higher spin theory.)
As evidence for the duality we have checked that the 
`wedge' subalgebra of $s{\cal W}_{\infty}$ is indeed ${\rm shs}(1|2)$ --- this is believed to be equivalent
to the statement that the asymptotic symmetry algebra of the Chern-Simons theory based on  ${\rm shs}(1|2)$ is 
$s{\cal W}_{\infty}$. We have also confirmed that the partition function of the ${\cal N}=1$ minimal models
is reproduced, in the 't~Hooft limit, by the thermal 1-loop partition function of the ${\rm shs}(1|2)$ higher spin theory
on AdS$_3$, where  in addition to the massless higher spin fields an ${\cal N}=1$ matter multiplet has been added. 

Our duality provides the first interesting example where the non-diagonal modular invariant of a 
${\cal W}_{\infty}$ theory has been identified with a dual higher spin theory. It would be very interesting
to study other non-diagonal modular invariants of ${\cal W}_{\infty}[\mu]$  (maybe for special values of
$\mu$), and see whether they also have interesting higher spin bulk duals.

\section*{Acknowledgements} 
We thank Rajesh Gopakumar for useful discussions. 
The work of MRG is supported in parts by the Swiss National Science Foundation. The paper is 
partially based on the Master thesis of one of us (MG). 

\appendix

%

\section{Supersymmetric form of branching functions}\label{app:B}

The Clebsch-Gordan coefficients for the mixed covariant-contravariant tensor representations of ${\rm U}(N)$, which are labeled by 
pairs of Young diagrams, have been computed in \cite{King:1971rs} (see also \cite{Naculich:1994nz})
\begin{equation}
c_{(\Pi_l,\Pi_r) \, (\Lambda_l,\Lambda_r)}^{(\alpha,\beta)}=\sum_{\pi,\xi,\omega,\gamma,\delta,\epsilon} c_{\pi \, \xi}^{\Lambda_l} \, c_{\omega \, \xi}^{\alpha} \, c_{\omega \, \gamma}^{\Pi_l} \, c_{\gamma \, \delta}^{\Lambda_r} \, c_{\delta \, \epsilon}^{\beta} \, c_{\pi \, \epsilon}^{\Pi_r}\ .
\label{mix}
\end{equation}
This was done with the help of an older formula
\begin{equation}
c_{(\Lambda_l,0)\, (0,\Lambda_r)}^{(\Xi_l,\Xi_r)}=\sum_\pi c_{\pi\,\Xi_l}^{\Lambda_l}c_{\pi\,\Xi_r}^{\Lambda_r}
\label{eq:original_king}
\end{equation}
derived in \cite{King:1970}, and the techniques developed in \cite{Abramsky:1970yd}.
The complicated nature of (\ref{mix}) is due to the cancellation between  boxes and antiboxes in the tensor 
product of $\Lambda=(\Lambda_l,\Lambda_r)$ with $\Pi=(\Pi_l,\Pi_r)$.
Indeed, for $|\pi|=0$ and $|\gamma|=0$ no boxes have canceled against anti-boxes, and eq. (\ref{mix}) factorises 
to $c_{\Pi_l \Lambda_l}^{\alpha} \, c_{\Pi_r \Lambda_r}^{\beta}$.
\smallskip

To illustrate the assembling of the branching function~\eqref{eq12} into the supersymmetric form~\eqref{eq:beauty},
let us consider first the simpler case where $\Lambda$ does not contain any anti-boxes, i.e.\ $\Lambda=(\Lambda_l,0)$.
Then eq.~\eqref{mix} simplifies to 
\begin{equation}
c_{(\Pi^t,\Pi) \, (\Lambda_l,0)^\ast}^{(\alpha,\beta)}
=\sum_{\gamma,\delta} c_{\gamma \, \alpha}^{\Pi^t}
c_{\gamma \, \delta}^{\Lambda_l} \, c_{\Pi \, \delta}^{\beta} \ , \label{oboxes} 
\end{equation}
where we recall the conjugation operation $(\Lambda_l,\Lambda_r)^\ast=(\Lambda_r,\Lambda_l)$
and note that the Clebsch-Gordon coefficient $c_{\pi\xi}^{0}$ is only non-zero (and equal to one) if
$\pi=\xi=0$. 
Plugging this relation into eq. (\ref{eq12}) leads to
\begin{align} \chi_{(\Lambda_l,0)}(q) 
&=q^{-\frac{c}{24}} \tilde{M}(q)  \, q^{-\frac{|\Lambda_l|}{4}} \sum_{\Pi,\gamma,\delta,\alpha} 
q^{\frac{|\Pi|}{2}} c_{\gamma \, \delta}^{\Lambda_l} \, c_{\gamma \, \alpha}^{\Pi^t} \, \textrm{ch}_{{\Pi}^t}(U_{\frac{1}{2}}) 
\cdot \textrm{ch}_{{\delta}^t}(U_{\frac{1}{2}}) \cdot \textrm{ch}_{{\alpha}^t}(U_{\frac{1}{2}}) \nonumber\\
&=q^{-\frac{c}{24}} \tilde{M}(q) \, q^{-\frac{|\Lambda_l|}{4}} \sum_{\gamma,\delta,\alpha} 
c_{\gamma \, \delta}^{\Lambda_l}\,
 \textrm{ch}_{\gamma}(U_1) \cdot \textrm{ch}_{{\delta}^t}(U_{\frac{1}{2}}) \cdot \textrm{ch}_{\alpha}(U_1) \cdot \textrm{ch}_{{\alpha}^t}(U_{\frac{1}{2}}) \nonumber\\
&= q^{-\frac{c}{24}} \tilde{M}(q) \sum_{\alpha} \textrm{ch}_{{\alpha}^t}(U_{\frac{3}{4}}) \cdot \textrm{ch}_{{\alpha}}(U_{\frac{3}{4}}) 
\sum_{\gamma,\delta}  c_{\gamma \, \delta}^{\Lambda_l^t} \, 
\textrm{ch}_{\delta}(U_{\frac{1}{4}}) \cdot \textrm{ch}_{\gamma^t}(U_{\frac{3}{4}}) \nonumber\\
&=\chi_0(q) \cdot \textrm{sch}_{\Lambda_l^t}(\mathcal{U}_{\frac{1}{4}})\ ,
\label{box}
\end{align}
where we have used the invariance of the Littlewood-Richardson coefficients under transposition, together with 
$|\Lambda_l|=|\pi|+|\xi|$. Furthermore, we have applied in the last step eq.~\eqref{eq9}, as well as 
\begin{equation}
\textrm{sch}_{\Lambda}(\mathcal{U}_{\frac{1}{4}})=\sum_{\gamma,\delta}  c_{\gamma \delta}^{\Lambda}\,
 \textrm{ch}_{\delta}(U_{\frac{1}{4}}) \cdot \textrm{ch}_{\gamma^t}(U_\frac{3}{4}) \label{sch}\ 
\end{equation}
that follows from \cite[eq.~(A.7) and (A.9)]{supersymmetric}. 

In order to treat the general case, we need two more character identities.
The first one is a generalisation of the Cauchy identity, see \cite[p.~93]{MacDonald} 
\begin{equation}
\sum_\rho \mathrm{ch}_{\rho/\lambda^t}(U_\frac{3}{4}) \cdot 
\mathrm{ch}_{\rho^t/\mu}(U_\frac{3}{4}) = \prod_{s=1}^\infty \prod_{n=s}^\infty (1+q^{n+\frac{1}{2}})\times 
\sum_\beta  \mathrm{ch}_{\lambda/\beta^t}(U_\frac{3}{4})\cdot \mathrm{ch}_{\mu^t/\beta}(U_\frac{3}{4})\ ,
\label{eq:smartid}
\end{equation}
where $\mathrm{ch}_{\lambda/\mu}=\sum_\nu c^\lambda_{\mu\nu}\mathrm{ch}_\nu$ are the skew Schur functions.
The second identity
\begin{equation}
\sum_\gamma \mathrm{ch}_{\lambda^t/\gamma^t}(U_\frac{1}{4})\cdot \mathrm{ch}_{\gamma/\beta}(U_\frac{3}{4})
=\sum_{\gamma,\epsilon,\delta} c^{\lambda^t}_{\gamma^t\beta^t}c^{\gamma^t}_{\epsilon\delta^t}\,
\mathrm{ch}_\epsilon(U_\frac{1}{4}) \cdot
\mathrm{ch}_\delta(U_\frac{3}{4})= \sum_{\gamma} c^{\lambda}_{\gamma\beta} \, 
\mathrm{sch}_{\gamma^t}(\mathcal{U}_\frac{1}{4})
\label{eq:id2}
\end{equation}
follows from the definition of the skew Schur functions, the identity (\ref{sch}), 
together with the 
associativity of the tensor product $\delta^t\otimes \beta^t\otimes \epsilon$ which implies that
\begin{equation}
\sum_\gamma c^{\lambda^t}_{\gamma^t\epsilon}\, c^{\gamma}_{\beta\delta}
=\sum_\gamma c^{\lambda^t}_{\gamma^t\epsilon}\, c^{\gamma^t}_{\beta^t\delta^t}
=\sum_\gamma c^{\lambda^t}_{\gamma^t\beta^t}\, c^{\gamma^t}_{\epsilon\delta^t} \ .
\end{equation}
Consider now the general case. Plugging eq.~(\ref{mix}) into eq.~(\ref{eq12}) and factorising characters we get
\begin{align}
\begin{split}
\chi_{(\Lambda_l,\Lambda_r)} &=  q^{-\frac{c}{24}} \tilde{M}(q) \, q^{-\frac{|\Lambda_l|+|\Lambda_r|}{4}}
\sum_{\Pi,\pi,\xi,\omega\atop{\gamma,\delta,\epsilon,\alpha,\beta}} \!\!\! q^{\frac{|\Pi|}{2}}
c_{\pi \, \xi}^{\Lambda_r}\, c_{\omega \, \gamma}^{\Pi^t}\, c_{\gamma \, \delta}^{\Lambda_l}\, c_{\pi \, 
\epsilon}^{\Pi}\, \textrm{ch}_{{\omega}^t}(U_{\frac{1}{2}}) \textrm{ch}_{{\xi}^t}(U_{\frac{1}{2}}) 
\textrm{ch}_{{\delta}^t}(U_{\frac{1}{2}}) \textrm{ch}_{{\epsilon}^t}(U_{\frac{1}{2}}) \end{split} \nonumber\\ 
\begin{split}
&= q^{-\frac{c}{24}} \tilde{M}(q) \, q^{-\frac{|\Lambda_l|+|\Lambda_r|}{4}} \sum_{\Pi,\pi,\gamma} q^{\frac{|\Pi|}{2}}
\textrm{ch}_{\Pi/\gamma^t}(U_{\frac{1}{2}})\,  \textrm{ch}_{{\Pi}^t/\pi^t}(U_{\frac{1}{2}})    \,
\textrm{ch}_{\Lambda_r^t/\pi^t}(U_{\frac{1}{2}})\,  \textrm{ch}_{\Lambda_l^t/\gamma^t}(U_{\frac{1}{2}})
\end{split}\nonumber\\
\begin{split}
&= q^{-\frac{c}{24}} \tilde{M}(q) \sum_{\Pi,\pi,\gamma} 
\textrm{ch}_{\Pi/\gamma^t}(U_{\frac{3}{4}}) \cdot \textrm{ch}_{{\Pi}^t/\pi^t}(U_{\frac{3}{4}})   \cdot
\textrm{ch}_{\Lambda_r^t/\pi^t}(U_{\frac{1}{4}}) \cdot \textrm{ch}_{\Lambda_l^t/\gamma^t}(U_{\frac{1}{4}})\ .
\end{split}\nonumber
\end{align}
Using now eq.~\eqref{eq:smartid} to perform the sum over $\Pi$ we obtain
\begin{align}\notag
\chi_{(\Lambda_l,\Lambda_r)}(q) &= \chi_0(q) 
\sum_{\beta,\pi,\gamma} 
\textrm{ch}_{\gamma/\beta^t}(U_{\frac{3}{4}})\cdot \textrm{ch}_{\pi/\beta}(U_{\frac{3}{4}}) \cdot 
\textrm{ch}_{\Lambda_r^t/\pi^t}(U_{\frac{1}{4}}) \cdot 
\textrm{ch}_{\Lambda_l^t/\gamma^t}(U_{\frac{1}{4}})\\ \notag
{}&= \chi_0(q) \sum_{\beta,\pi,\gamma} c^{\Lambda_l}_{\gamma\,\beta^t}\,c^{\Lambda_r}_{\beta\,\pi}\,
\mathrm{sch}_{\gamma^t}(\mathcal{U}_{\frac{1}{4}}) \cdot \mathrm{sch}_{\pi^t}(\mathcal{U}_{\frac{1}{4}})\\
{}&= \chi_0(q) \sum_{\gamma,\pi} c_{(\Lambda^t_l,0)\,(0,\Lambda_r)}^{(\gamma^t,\pi)} \,
\mathrm{sch}_{\gamma^t}(\mathcal{U}_{\frac{1}{4}})\cdot \mathrm{sch}_{\pi^t}(\mathcal{U}_{\frac{1}{4}})\ ,
\label{sub}
\end{align}
where in the second step we have used eq.~\eqref{eq:id2}, and in the last step eq.~\eqref{eq:original_king}.
Note that for $\Lambda=(\Lambda_l,0)$, i.e.\ $\Lambda_r=0$, the result indeed reduces to eq.~(\ref{box}).

\section{Explicit constraints on the OPE coefficients from associativity}

In this appendix, we give more details about the results of studying the associativity constraints. 
The normalisation of higher spin composites can be fixed by comparing with the leading 
terms listed in Appendix~\ref{app:composites}.
\smallskip

\noindent {\bf B.1 \ The OPE $\frac{7}{2}\times \frac{9}{2}$} 
\medskip

\noindent 
The ansatz for the OPE was given in (\ref{OPE79}). The associativity of $\super 5 \times \super 7 \times \super 7$ 
leads to the constraints that the couplings take the values
\begin{align}
\coupling{\frac{7}{2}}{\frac{9}{2}}{\frac{5}{2}} &= \frac{28800 (2 c+5) (6 c-13) (7 c-10)}{c^2 (2 c+37) (10 c-7)^2}\ ,\\
\coupling{\frac{7}{2}}{\frac{9}{2}}{\frac{9}{2}} &=\frac{192 \left(110 c^3+881 c^2-3439 c-8766\right)}{c (2 c+37) 
(4c+21) (10 c-7)} \ , \\
\coupling{\frac{7}{2}}{\frac{9}{2}}{6,b} &= \frac{8400 (6 c-13)}{c (2 c+37) (10 c-7)}\ ,\\
\coupling{\frac{7}{2}}{\frac{9}{2}}{\frac{13}{2},a} &= 
\frac{35 (20 c+283)}{624 (2 c+53)} \coupling{\frac{5}{2}}{\frac{9}{2}}{\frac{11}{2},a}+\frac{5}{6}
\coupling{\frac{5}{2}}{\frac{11}{2}}{\frac{13}{2},a}+\frac{700 (6 c-13) (2345 c+26774)}{143 c
 (2 c+37) (2 c+53) (10 c-7)}\ ,\\
\coupling{\frac{7}{2}}{\frac{9}{2}}{7,a} &= \frac{1200 (94 c+375)}{13 c (2 c+37) (10 c-7)}\ ,\\
\coupling{\frac{7}{2}}{\frac{9}{2}}{\frac{15}{2},c} &= -\frac{86400 (c+20) (6 c-13)}{7 c (2 c+37) (10 c-7) (13 c+162)}\ .
\end{align}
Furthermore, the other couplings vanish
\be
\coupling{\frac{7}{2}}{\frac{9}{2}}{\frac{11}{2}} =
\coupling{\frac{7}{2}}{\frac{9}{2}}{\frac{11}{2},a} =
\coupling{\frac{7}{2}}{\frac{9}{2}}{6,a} =
\coupling{\frac{7}{2}}{\frac{9}{2}}{7,b} =
\coupling{\frac{7}{2}}{\frac{9}{2}}{\frac{15}{2}} =
\coupling{\frac{7}{2}}{\frac{9}{2}}{\frac{15}{2},a} =
\coupling{\frac{7}{2}}{\frac{9}{2}}{\frac{15}{2},b} =
\coupling{\frac{7}{2}}{\frac{9}{2}}{\frac{15}{2},d} =0\ .
\ee
\medskip

\noindent {\bf B.2 \ The OPE $\frac{5}{2}\times \frac{11}{2}$}
\smallskip

\noindent The ansatz for the OPE was given in (\ref{OPE511}). The associativity of $\super 5 \times \super 7 \times \super 7$ 
leads to the constraints that the couplings take the values
\begin{align}
\coupling{\frac{5}{2}}{\frac{11}{2}}{\frac{5}{2}} &= \frac{34560 (2 c+5) (6 c-13) (7 c-10) (17 c-24)}{11 c^2 (2 c+37)
   (2 c+53) (10 c-7)^2}\nonumber \\
   & \qquad\qquad -\frac{4 (2 c+5) \left(20 c^3+1441   c^2+7180 c+2384\right)}{5 c (c+11) (2 c+53)
   (10 c-7)} \,\coupling{\frac{5}{2}}{\frac{9}{2}}{\frac{11}{2},a}\ ,\\
\coupling{\frac{5}{2}}{\frac{11}{2}}{\frac{9}{2}} &= \frac{576 \left(1210 c^3+17801 c^2+50654 c-327272\right)}{55 c 
(2 c+37) (2 c+53) (10 c-7)}-\frac{3 (17 c-24)}{25 (2 c+53)}\, \coupling{\frac{5}{2}}{\frac{9}{2}}{\frac{11}{2},a}\ ,\\
\coupling{\frac{5}{2}}{\frac{11}{2}}{6,b} &= \frac{7 (4 c+171)}{10 (2 c+53)}\, \coupling{\frac{5}{2}}{\frac{9}{2}}{\frac{11}{2},a}
-\frac{10080 (6  c-13) (23 c-11)}{11 c (2 c+37) (2 c+53) (10 c-7)}\ , \\
\coupling{\frac{5}{2}}{\frac{11}{2}}{\frac{13}{2}} &= \frac{6}{5}\ ,\\
\coupling{\frac{5}{2}}{\frac{11}{2}}{7,a} &= \frac{(4 c-11)}{26 (2 c+53)}\, \coupling{\frac{5}{2}}{\frac{9}{2}}{\frac{11}{2},a}
+\frac{12960 \left(334 c^2+4983 c+4224\right)}{143 c (2 c+37) (2 c+53) (10   c-7)} \ , \\
\coupling{\frac{5}{2}}{\frac{11}{2}}{\frac{15}{2},c} &= \frac{86400 (c+20) (6 c-13) (23 c-11)}{77 c (2 c+37) (2 c+53) (10
   c-7) (13 c+162)}-\frac{6 (c+20) (4 c+171) }{7 (2 c+53) (13 c+162)}\,\coupling{\frac{5}{2}}{\frac{9}{2}}{\frac{11}{2},a}\ .
\end{align}
Furthermore, the remaining couplings vanish
\be
\coupling{\frac{5}{2}}{\frac{11}{2}}{\frac{7}{2}} =
\coupling{\frac{5}{2}}{\frac{11}{2}}{\frac{11}{2},a} =
\coupling{\frac{5}{2}}{\frac{11}{2}}{6,a} =
\coupling{\frac{5}{2}}{\frac{11}{2}}{7,b} =
\coupling{\frac{5}{2}}{\frac{11}{2}}{\frac{15}{2}} =
\coupling{\frac{5}{2}}{\frac{11}{2}}{\frac{15}{2},a} =
\coupling{\frac{5}{2}}{\frac{11}{2}}{\frac{15}{2},b} =
\coupling{\frac{5}{2}}{\frac{11}{2}}{\frac{15}{2},d} =0\ .
\ee
\medskip

\noindent {\bf B.3 \ The OPE $\frac{7}{2}\times \frac{11}{2}$}
\smallskip


\noindent The ansatz for the OPE $\super 7 \times \super{11}$ is 
\begin{align}
\super 7 \times \super{11} &= 
\coupling{\frac{7}{2}}{\frac{11}{2}}{\frac{5}{2}}\,\super 5+
\coupling{\frac{7}{2}}{\frac{11}{2}}{\frac{7}{2}}\,\super 7+
\coupling{\frac{7}{2}}{\frac{11}{2}}{\frac{9}{2}}\,\super 9+
\coupling{\frac{7}{2}}{\frac{11}{2}}{\frac{11}{2}}\,\super{11}+
\coupling{\frac{7}{2}}{\frac{11}{2}}{\frac{11}{2},a}\,\extrasuper{11}{a} \nonumber\\
&+\sum_{I=a,b}\coupling{\frac{7}{2}}{\frac{11}{2}}{6,I}\,\extracompsuper{6}{I}
+\coupling{\frac{7}{2}}{\frac{11}{2}}{\frac{13}{2}}\,\super{13}+
\coupling{\frac{7}{2}}{\frac{11}{2}}{\frac{13}{2},a}\,\extrasuper{13}{a}+
\sum_{I=a,b}\coupling{\frac{7}{2}}{\frac{11}{2}}{7,I}\,\extracompsuper{7}{I}\nonumber \\
&
+\coupling{\frac{7}{2}}{\frac{11}{2}}{\frac{15}{2}}\,\super{15}+
\sum_{I=a, b, c}\coupling{\frac{7}{2}}{\frac{11}{2}}{\frac{15}{2},I}\,\extrasuper{15}{I}\nonumber \\
&
+\sum_{I=a, \dots, f}\coupling{\frac{7}{2}}{\frac{11}{2}}{8,I}\,\extracompsuper{8}{I}+
\coupling{\frac{7}{2}}{\frac{11}{2}}{ \frac{17}{2}}\,\super{17}+
\sum_{I=a, \dots, e}\coupling{\frac{7}{2}}{\frac{11}{2}}{8,I}\,\extrasuper{17}{I}.
\end{align}
Imposing the   associativity of $\super 7 \times \super 7 \times \super 7$, the 
various couplings turn out to equal
\begin{align}
\coupling{\frac{7}{2}}{\frac{11}{2}}{\frac{7}{2}} &=  \frac{13824 (4 c+21) (6 c-13) \left(605 c^2+1278 c-8456\right)}{11 c^2 
(2 c+37) (2c+53) (10 c-7)^2}\ ,  \label{eq:7117}\\
\coupling{\frac{7}{2}}{\frac{11}{2}}{\frac{11}{2}} &=  \frac{336 \left(2288 c^4+47930 c^3+189489 c^2-2153000
   c-4668092\right)}{11 c (2 c+37) (2 c+53) (4 c+21) (10 c-7)}\ ,\\
\coupling{\frac{7}{2}}{\frac{11}{2}}{6,a} &=  -\frac{829440 (14 c+11) \left(502 c^2+8203 c-28424\right)}{121 c^2 (2 c+37) (2 c+53) (10
   c-7)^2}\ ,\\
\coupling{\frac{7}{2}}{\frac{11}{2}}{8,a} &=  \mbox{complicated rational function of $c$}\ , \\
\coupling{\frac{7}{2}}{\frac{11}{2}}{\frac{11}{2},a} &=  \frac{46448640 (c+11) (6 c-13) (7 c-10) (23 c-11)}{121
   c^2 (2 c+37)^2 (2 c+53) (10 c-7)^2}\ ,\\
\coupling{\frac{7}{2}}{\frac{11}{2}}{7,b} &=  \frac{20736 (26 c-77)}{11 c (2 c+37) (10 c-7)}\ ,\\
\coupling{\frac{7}{2}}{\frac{11}{2}}{8,d} &= 
   \frac{64 \left(58924 c^2+963796 c+2518087\right)}{55 c (2 c+37) (2 c+53) (10 c-7) (13 c+5)}\ ,\\
   \coupling{\frac{7}{2}}{\frac{11}{2}}{\frac{17}{2},d} &= 
   \frac{24300 (26 c-77)}{11 c (2 c+37) (5 c+77) (10 c-7)}\ ,\\
   \coupling{\frac{7}{2}}{\frac{11}{2}}{8,e} &=  -\frac{32 \left(5024 c^3+266906
   c^2+1815131 c+1654059\right)}{55 c (2 c+37) (2 c+53) (10 c-7) (13 c+5)}\ .
\end{align}
Furthermore, the other couplings vanish
\begin{align}
\coupling{\frac{7}{2}}{\frac{11}{2}}{\frac{5}{2}} =
\coupling{\frac{7}{2}}{\frac{11}{2}}{\frac{9}{2}} =
\coupling{\frac{7}{2}}{\frac{11}{2}}{\frac{13}{2}} =
\coupling{\frac{7}{2}}{\frac{11}{2}}{\frac{17}{2}} =
\coupling{\frac{7}{2}}{\frac{11}{2}}{7,a} =
\coupling{\frac{7}{2}}{\frac{11}{2}}{\frac{13}{2},a} =
\coupling{\frac{7}{2}}{\frac{11}{2}}{\frac{17}{2},a} =
\coupling{\frac{7}{2}}{\frac{11}{2}}{6,b} \nonumber  \\
= \coupling{\frac{7}{2}}{\frac{11}{2}}{8,b} =
\coupling{\frac{7}{2}}{\frac{11}{2}}{\frac{17}{2},b} =
\coupling{\frac{7}{2}}{\frac{11}{2}}{8,c} =
\coupling{\frac{7}{2}}{\frac{11}{2}}{\frac{17}{2},c} =
   \coupling{\frac{7}{2}}{\frac{11}{2}}{\frac{17}{2},e} =
   \coupling{\frac{7}{2}}{\frac{11}{2}}{8,f} =0\ .
\end{align}
The choice of basis for the various composite superprimaries is largely arbitrary. We determined them automatically,
and as a consequence some of the coupling constants are 
complicated rational functions, e.g.\  the one appearing in $\coupling{\frac{7}{2}}{\frac{11}{2}}{8,a}$. 
\medskip

\noindent {\bf B.4 \ The OPEs $\frac{9}{2}\times\frac{9}{2}$ and  $\frac{5}{2}\times\frac{13}{2}$}
\smallskip


\noindent 
We have also made the most general ansatz for the OPEs
$\super 9\times \super 9$ and $\super 5\times \super{13}$
\begin{align}
\super 9 \times \super{9} &=
\coupling{\frac{9}{2}}{\frac{9}{2}}{0} \mathbb I + 
\coupling{\frac{9}{2}}{\frac{9}{2}}{\frac{7}{2}}\,\super 7+
\coupling{\frac{9}{2}}{\frac{9}{2}}{\frac{11}{2}}\,\super{11}+
\coupling{\frac{9}{2}}{\frac{9}{2}}{\frac{11}{2},a}\,\extrasuper{11,a}+
\sum_{I=a,b}\coupling{\frac{9}{2}}{\frac{9}{2}}{6,I}\,\extracompsuper{6}{I}\nonumber \\
&
+\coupling{\frac{9}{2}}{\frac{9}{2}}{\frac{15}{2}}\,\super{15}+
\sum_{I=a, b, c}\coupling{\frac{9}{2}}{\frac{9}{2}}{ \frac{15}{2},I}\,\extrasuper{15}{I}+
\sum_{I=a, \dots, f}\coupling{\frac{9}{2}}{\frac{9}{2}}{8,I}\,\extracompsuper{8}{I}\ , \\
\super 5 \times \super{13} &=
\coupling{\frac{5}{2}}{\frac{13}{2}}{\frac{7}{2}}\,\super 7+
\coupling{\frac{5}{2}}{\frac{13}{2}}{\frac{9}{2}}\,\super 9+
\coupling{\frac{5}{2}}{\frac{13}{2}}{\frac{11}{2}}\,\super{11}+
\coupling{\frac{5}{2}}{\frac{13}{2}}{\frac{11}{2},a}\,\extrasuper{11}{a}+
\sum_{I=a,b}\coupling{\frac{5}{2}}{\frac{13}{2}}{6,I}\,\extracompsuper{6}{I}\nonumber \\
&
+\sum_{I=a,b}\coupling{\frac{5}{2}}{\frac{13}{2}}{7,I}\,\extracompsuper{7}{I}+
\coupling{\frac{5}{2}}{\frac{13}{2}}{\frac{15}{2}}\,\super{15}+
\sum_{I=a, b, c}\coupling{\frac{5}{2}}{\frac{13}{2}}{ \frac{15}{2},I}\,\extrasuper{15}{I}\nonumber \\
&
+\sum_{I=a, \dots, f}\coupling{\frac{5}{2}}{\frac{13}{2}}{ 8,I}\,\extracompsuper{8}{I}+
\coupling{\frac{5}{2}}{\frac{13}{2}}{ \frac{17}{2}}\,\super{17}+
\sum_{I=a, \dots, e}\coupling{\frac{5}{2}}{\frac{13}{2}}{ 8,I}\,\extrasuper{17}{I}\ .
\end{align}
The associativity of
$\super 5\times \super 7\times \super 9$ and $\super 5\times \super 5\times \super{11}$ can then 
be imposed as in the previous steps. We do not write the detailed list of couplings that are fixed in this way,
except for giving the specific case of 
\be\label{eq:990}
\coupling{\frac{9}{2}}{\frac{9}{2}}{0}=
\frac{28800 (2 c+5) (6 c-13) (7 c-10)}{(7-10 c)^2 c^2 (2 c+37)}\ ,
\ee
that will play a role for the truncation analysis of Section~\ref{sec:trunc}.

\section{Normalisation of the composite superprimaries}
\label{app:composites}

In order to fix our conventions, we report here the leading contributions to the composite superprimaries; 
the remaining terms that are needed in order to make them superprimaries are 
$G$ and $T$ descendants of fields of lower conformal dimension. We also only specify the `top' component
of the superprimary. For $\extrasuper{11}{a}$, $\mb V^{(6), a}$, 
$\mb V^{(6), b}$ and $\extrasuper{13}{a}$ the relevant expressions
were already given after (\ref{eq:first-composites}) and in (\ref{eq:second-composite}). The other cases of interest to us are 
\begin{align}
V^{(7), a} &= \frac{1}{5}\,(V^{(3)}V^{(4)})+\cdots\ ,\\
V^{(7),b} &= (V^{(\frac{5}{2})}V^{(\frac{9}{2})})+\cdots\ , \\
V^{(\frac{15}{2}),a} &= -\frac{\left(2 c^2+151 c+2166\right) (V^{(\frac{5}{2})}{V^{(3)}}'')}{6 (25 c+956)}+\frac{\left(14 c^2+673 c+6510\right) ({V^{(\frac{5}{2})}}'{V^{(3)}}')}{15 (25 c+956)}\nonumber \\
& -\frac{\left(14 c^2+757
   c+3690\right) ({V^{(\frac{5}{2})}}''V^{(3)})}{30 (25 c+956)}\nonumber \\
   & +\frac{\left(240 c^4+31436 c^3+1137640 c^2+13454979 c+21528990\right)
   {V^{(\frac{7}{2})}}''''}{960 (25 c+956) \left(20 c^2+708 c+1197\right)}+\cdots\ , \\
V^{(\frac{15}{2}),b} &= (V^{(\frac{7}{2})}V^{(4)})+\cdots, \\
V^{(\frac{15}{2}),c} &= \frac{2 (4 c+31) (V^{(3)}{V^{(\frac{7}{2})}}')}{5 (c+20)}-\frac{2 (4 c+31) ({V^{(\frac{5}{2})}}'V^{(4)})}{5 
(c+20)}-\frac{14 (2 c+33)
   ({V^{(3)}}'V^{(\frac{7}{2})})}{15 (c+20)}\nonumber \\
   & +(V^{(\frac{5}{2})}{V^{(4)}}')-\frac{32
   \left(10 c^4-421 c^3-3317 c^2+26691 c+10112\right) {V^{(\frac{5}{2})}}'''''}{25 c (c+20) (2 c+37) (4 c+21) (10 c-7)}\nonumber \\
   & -\frac{(2 c+19){V^{(\frac{9}{2})}}'''}{15 (c+20)}+\cdots\ , \\
V^{(\frac{15}{2}),d} &= (V^{(\frac{5}{2})}V^{(5)})+\frac{9}{5}(V^{(3)}V^{(\frac{9}{2})})\cdots\ . 
\end{align}

\end{document}